\documentclass[a4paper,fleqn,usenatbib]{mnras}

\usepackage[T1]{fontenc}
\usepackage{aecompl}
\usepackage{graphicx}	
\usepackage{amsmath}	
\usepackage{amssymb}	
\usepackage{bm}		
\usepackage{makecell} 
\usepackage{array}	

\DeclareMathOperator{\erf}{erf} 

\title[Is the VPOS a serious problem for $\Lambda$CDM?]{Is the vast polar structure of dwarf galaxies a serious problem for $\Lambda$ cold dark matter?}

\author[A. Lipnicky and S. Chakrabarti]{
Andrew Lipnicky$^{1}$\thanks{E-mail: awl6964@rit.edu (AL)}
and Sukanya Chakrabarti$^{1}$
\\
$^{1}$School of Physics and Astronomy, Rochester Institute of Technology, 1 Lomb Memorial Dr, Rochester, NY 14623, USA
}

\date{Accepted 2017 January 31. Received 2016 December 21; in original form 2016 July 13}

\pubyear{2016}

\begin{document}
\label{firstpage}
\pagerange{\pageref{firstpage}--\pageref{lastpage}}
\maketitle

\begin{abstract}
The dwarf galaxies around the Milky Way are distributed in a so-called vast polar structure (VPOS) that may be in conflict with $\Lambda$ cold dark matter ($\Lambda$CDM) simulations. Here, we seek to determine if the VPOS poses a serious challenge to the $\Lambda$CDM paradigm on galactic scales. Specifically, we investigate if the VPOS remains coherent as a function of time. Using the measured \textit{Hubble Space Telescope (HST)} proper motions and associated uncertainties, we integrate the orbits of the classical Milky Way satellites backwards in time and find that the structure disperses well before a dynamical time. We also examine in particular Leo I and Leo II using their most recent proper motion data, both of which have extreme kinematic properties, but these satellites do not appear to drive the polar fit that is seen at the present day. We have studied the effect of the uncertainties on the \textit{HST} proper motions on the coherence of the VPOS as a function of time. We find that 8 of the 11 classical dwarfs have reliable proper motions; for these eight, the VPOS also loses significance in less than a dynamical time, indicating that the VPOS is not a dynamically stable structure. Obtaining more accurate proper motion measurements of Ursa Minor, Sculptor, and Carina would bolster these conclusions. 
\end{abstract}




\section{Introduction}

A topic of recent interest and controversy is that concerning planes of dwarf galaxies in the Local Group. It has long been understood that an apparent plane of dwarf galaxies resides around the Milky Way very near the galactic poles, deemed the Vast Polar Structure \citep[VPOS,][]{Kunkel:1976aa,Lynden-Bell:1976aa, Kroupa:2005aa, Zentner:2005aa, Pawlowski:2013aa}, and recent work \citep{Conn:2013aa,Ibata:2013aa} have also found a thin distribution of corotating galaxies around M31. Work by \citet{Shaya:2013aa} found that 43 of the 50 Local Group satellites within 1.1 Mpc are contained within four different planes. The observations of these planar structures appear to challenge the current $\Lambda$ cold dark matter ($\Lambda$CDM) theory of hierarchical structure formation and call into question its validity (Kroupa et al. 2005; Pawlowski et al. 2015). 

 Following the initial impression of the VPOS as a challenge to $\Lambda$CDM, various authors have examined the effects of large-scale structure and baryonic physics on the VPOS, and have investigated its statistical significance relative to cosmological simulations that in fact do not manifest purely isotropic distributions, as assumed earlier by \citet{Kroupa:2005aa}. \citet{Zentner:2005aa} showed first that an isotropic distribution of sub-haloes is not the correct null hypothesis for testing $\Lambda$CDM (even for dissipationless simulations), and showed that the origin of the flattening may be due to the preferential accretion of satellites along the major axis of the halo. Further insight into this problem was obtained by works such as \citet{Shaya:2013aa} and \citet{McCall:2014aa}, which have examined how the cosmic web affects sub-galactic structures. \citet{Shaya:2013aa} found that the planar structure of the Local Group dwarfs is consistent with large-scale structure and due to the evacuation of the Local Void. Similarly, \citet{Libeskind:2015aa} showed that the alignment of the Local Group dwarfs along the shear field agrees with the $\Lambda$CDM paradigm. \citet{Wang:2013aa} have shown that in terms of spatial distribution, the Milky Way satellite plane is only in the 5--10 per cent tail of this distribution of planes found in simulations. \citet{Cautun:2015aa} found a similar solution for both the Milky Way and M31 planes and state that the rarity of the observed planes is largely due to a posteriori defined tests and misinterpretation of results. The analysis of the VPOS in dissipationless cosmological simulations may be summarized as follows -- these simulations predict sub-haloes to be distributed anisotropically \citep{ Wang:2013aa, Cautun:2015aa}; however, the \emph{degree} to which the VPOS is arranged seems to be at odds with $\Lambda$CDM \citep[e.g.,][]{Pawlowski:2014ab}. It is of note that the degree of anisotropy is especially pronounced for the more massive satellites (Libeskind et al. 2014). 

Various authors have also examined the effects of baryonic physics as a possible solution to this problem \citep[][and others]{Libeskind:2007aa, Sawala:2014aa}. \citet{Sawala:2014aa, Sawala:2016aa} argued that many of the discrepancies between the $\Lambda$CDM paradigm and observations on sub-galactic scales, such as the Missing Satellites problem \citep{Klypin:1999aa, Moore:1999aa, Kravtsov:2004aa} and the Too Big To Fail problem \citep{Boylan-Kolchin:2011aa}, can be resolved using hydrodynamical cosmological simulations that are designed to match the Local Group environment and may even play a part in the Planes of Satellites problem. However, \cite{Pawlowski:2015aa} have argued that these resolutions may be problematic. In particular, they find that \citet{Sawala:2014aa}'s resolution of the VPOS problem is due to ignoring the radial positions of the satellites. Moreover, although \citeauthor{Sawala:2014aa} performed a hydrodynamical simulation, they did not contrast their results with a dark matter-only simulation, as has been done recently by \citet{Ahmed:2016aa}, who find that the inclusion of baryons significantly changes the radial distribution of the satellites, thereby increasing the significance of planar structures. 

 It has also been suggested that these planes of dwarf galaxies are tidal dwarfs, having been pulled from the Milky Way during a past interaction with M31 \citep{Kroupa:2005aa, Hammer:2013aa, Pawlowski:2013aa}. Therefore, one might expect the planar dwarf galaxies to be distinct from off-plane dwarfs due to their different formation histories. However, \citet{Collins:2015aa} found that there is no difference in the observed properties (sizes, luminosities, masses, velocities, metallicities and star formation histories) for the on and off plane dwarf galaxies in M31. 
 
Yet another class of solutions lies in the as yet unknown form of the dark matter particle. Solutions which use dissipative dark matter physics may be able to explain planar structures and solve other issuers such as the Missing Satellites problem but full simulations have yet to be performed \citep{Randall:2015aa, Foot:2016aa}.

 What we seek to do here is to investigate the stability of the VPOS and the membership of the VPOS using the recently obtained \textit{Hubble Space Telescope (HST)} proper motions of the classical Milky Way satellites. If the VPOS is a serious problem for $\Lambda$CDM, one expects that it should persist over a dynamical time and should not be unique to the present day. Dynamical coherence over long time-scales would presumably occur for satellites with aligned angular momentum vectors, as has been claimed by prior work \citep{Pawlowski:2014aa}, and thus it is critical to examine the long-term stability of this structure. Moreover, there may be certain satellites that drive the appearance of the planar structure at the present day. If so, it is critical to examine whether a subset excluding these satellites resembles cosmological simulations.

A significant recent advance for near-field cosmology are the proper motions that have been obtained using \textit{HST} for all of the classical Milky Way satellites \citep{Kallivayalil:2013aa, Sohn:2013aa}. These measurements now enable us to calculate realistic orbits for these satellites. Although there are errors associated with the proper motion measurements, several authors have been able to make substantive new inferences. For example, \citet{Kallivayalil:2013aa} found using the third-epoch \textit{HST} Magellanic Cloud proper motion measurements that periods less than 4 Gyr are ruled out, which is a challenge for traditional Magellanic Stream models. Furthermore, \citet{Kallivayalil:2013aa} found that if one assumes that the Large Magellanic Cloud (LMC) and Small Magellanic Cloud (SMC) have been a bound pair for a few Gyr, then first infall models are preferred. These proper motions can also inform our understanding of the VPOS -- both its stability and the membership of the VPOS. When considering a possible plane of satellites, it is imperative to determine which satellites truly belong in such an analysis. In early studies of the VPOS, only position information was available to the authors \citep[e.g.][]{Kroupa:2005aa, Zentner:2005aa}; however, we now have accurate three-dimensional (3D) space and velocity information, which enables one to look at the orbital elements \citep{Pawlowski:2013aa} and the past orbital history of each member. Much work has been done investigating the previous orbits of these satellites in order to constrain the mass of the Milky Way, the shape, and the extent of the Milky Way's dark matter halo. For example, \citet{Sohn:2013aa} and \citet{Boylan-Kolchin:2013aa} used the motion of Leo I to constrain the virial mass and extent of the Milky Way. 

This paper is organized as follows. In \S 2, we discuss the dissipationless cosmological simulations that were used to compare to the Milky Way sample of dwarf galaxies, our method of plane fitting and the statistic measures used to compare our samples. We also discuss the orbit calculations we performed to analyse the longevity of the VPOS. In \S 3, we investigate individual members of the VPOS and explore the results of looking at subsets of the classical dwarfs. In \S 4, we discuss the results of the orbit calculations, the effects of the errors, and limitations of our analysis. Finally, in \S 5, we discuss the impact of these results, and conclude in \S 6.

\section{Methods}

The satellites that are used in our analysis are considered the `classical' dwarf satellites and are the only satellites for which we have reliable proper motion measurements. The position and velocity vectors for the 11 classical satellites are given in Table \ref{dwarf_data}.

\subsection{Dissipationless cosmological simulations}
Following the work of earlier groups \citep{Kroupa:2005aa,Zentner:2005aa}, we compare the Milky Way distribution of satellites to current dark matter simulations of Milky Way-like haloes. 

Our first comparison is to Via Lactea II \citep[VLII;][]{Diemand:2007aa}, where an $N$-body code was used to model a Milky Way-sized halo to the present epoch ($z=0$) using over 200 million particles. The simulations were performed with \textsc{PDKGRAV} \citep{Stadel:2001aa, Wadsley:2004aa} and adopted the best-fitting cosmological parameters from the \textit{Wilkinson Microwave Anisotropy Probe (WMAP)} three year data release \citep{Spergel:2007aa}: $\Omega_\text{M} = 0.238$, $\Omega_{\Lambda}=0.762$, $H_0 = 73$ km s$^{-1}$ Mpc$^{-1}$, $n = 0.951$, and $\sigma_8 = 0.74$. The host halo suffered no major mergers after $z=1.7$ and had a host halo mass of $M_{\text{halo}}=1.77\times10^{12} \text{M}_{\odot}$ within a radius of $r_{\text{vir}}= 389$ kpc, making it a good candidate for a Milky Way-like disc galaxy at the present day. VLII has a sub-halo mass resolution limit of $M_{\text{sub}}=4\times10^6 \text{M}_{\odot}$ and is therefore capable of resolving all the classical Milky Way dwarfs.

We also compare the Milky Way distribution to the simulations performed by \citet{Garrison-Kimmel:2014aa}: Exploring the Local Volume in Simulations (ELVIS). The ELVIS simulations were performed using \textsc{GADGET-3} and \textsc{GADGET-2} \citep{Springel:2005aa}, both of which are tree-SPH codes that follow the dissipationless component with the $N$-body method. ELVIS is a dissipationless cosmological simulation with adopted $\Lambda$CDM parameters from \textit{WMAP}-7 \citep{Larson:2011aa}: $\Omega_\text{M} = 0.266$, $\Omega_{\Lambda}=0.734$, $H_0 = 71$ km s$^{-1}$ Mpc$^{-1}$, $n = 0.963$, and $\sigma_8 = 0.801$. Simulations were chosen to be good analogues of the Local Group and had to meet a specific set of criteria based on host mass, total mass, separation, radial velocity, and isolation. In total, 12 pairs of galaxies were simulated to model the Milky Way -- M31 system; these 24 haloes were then simulated again in isolation. The isolated simulations were shown to have similar subhalo counts and mass functions. However, sub-haloes in paired simulations were shown to have substantially higher tangential velocities. For the comparisons made in this paper, we choose to compare the Milky Way dwarf population to the high resolution, isolated simulations referred to as iScylla and iHall. These two simulations have a subhalo mass resolution of $M_{\text{sub}}\sim2\times10^5 \text{M}_{\odot}$, while all of the other ELVIS simulations have a mass resolution of $M_{\text{sub}}>2\times10^7 \text{M}_{\odot}$ and therefore are not capable of resolving all the classical Milky Way dwarfs. These haloes have similar properties to the Milky Way and are good analogues. The properties of the realized haloes from both simulations are given in Table \ref{DM_sim_values}. 

\begin{table*}
\caption{Summary of simulation values compared to the Milky Way. The last column refers to the number of sub-haloes that have $V_\text{peak} > 10$ km s$^{-1}$ and are within 300 kpc of the Milky Way. The virial mass of the Milky Way is taken from \citet{Boylan-Kolchin:2013aa}, and the number of sub-haloes of the Milky Way was gathered by \citet{McConnachie:2012aa}. }
\begin{center}
\begin{tabular}{ccccc} \hline\hline
Halo			& $M_{\text{vir}} (10^{12} \text{M}_{\odot})$	& $r_{\text{vir}}$ (kpc)&$V_{\text{vir}}$ (km s$^{-1}$)& $n_{\text{haloes}}$ \\ \hline
Via Lactea II	& 1.77						&	389		&	140		& 874 	\\ 
iScylla HiRes	& 1.61						&	305		&	150		& 420 \\ 
iHall	HiRes	& 1.67						&	309		&	130		& 604 \\
Milky Way		& $1.6^{+0.8}_{-0.6}$	&	304		&	150		& $\geq $27	\\ \hline
\end{tabular}
\end{center}
\label{DM_sim_values}
\end{table*}

\subsection{Plane fitting: the dwarf galaxies at the present day}

The method used to fit planes to distributions of dwarf galaxies in both simulations and the Milky Way is the method of principal component analysis (PCA). The best-fitting plane is $\hat{\bm{n}}\bullet\bm{x}=\hat{n}_xX+\hat{n}_yY+\hat{n}_zZ = 0$, where $\hat{\bm{n}}$ is the normal vector of the best-fitting plane and $X, Y, Z$ are the coordinate points for a satellite. Note that our solution forces a best-fitting plane to go through the centre since an off-centre solution would not be meaningful. To perform PCA, we evaluate the covariance matrix and perform an eigenvalue analysis. The resulting eigenvector associated with the smallest eigenvalue is the normal of the plane, $\hat{\bm{n}}$, which passes through the origin and ensures that the found plane is the best solution. In other words, a vector pointing along the direction of least variance will be perpendicular to the direction of greatest variance and therefore represents the normal vector of a best-fitting plane.

The distance that the $i$-th dwarf out of $N$ dwarfs lies above the plane is determined, and the total rms distance of the distribution is calculated as a measure of planarity: 

\begin{equation}
D_{\text{rms}} = \sqrt{\frac{1}{N}\sum_i^N (\hat{\bm{n}}\bullet{\bm{x}_i})^2}.
\label{D_rms}
\end{equation}

 However, this comparison is not sufficient to give a full understanding of the distribution as it does not take into account the compactness of the distributions as a more compact distribution will naturally lead to a small $D_{\text{rms}}$ regardless of the actual distribution \citep{Kang:2005aa, Zentner:2005aa, Metz:2007aa}. This effect can be nullified by normalizing $D_{\text{rms}}$ by the median radial distance, $R_{\text{med}}$. We define this measure of compactness as \citet{Zentner:2005aa} have defined:

\begin{equation}
\delta = \frac{D_{\text{rms}}}{R_{\text{med}}}.
\label{delta}
\end{equation}

\begin{figure}
\begin{center}
\includegraphics[width=\columnwidth]{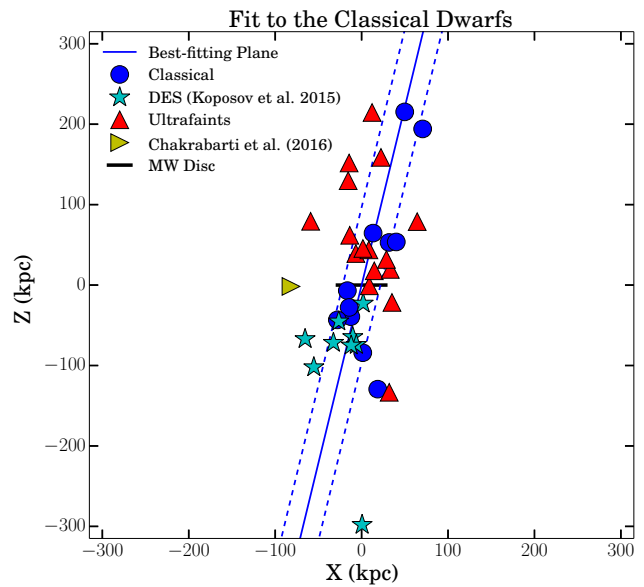}
\caption{All known dwarf galaxies surrounding the Milky Way are displayed (including those that are not spectroscopically confirmed) and the VPOS is shown via the solid blue line. The solid horizontal line in the centre represents the Milky Way galactic disc, and the dotted lines bordering the VPOS represent the rms distance, $D_{\text{rms}}=21.3$ kpc, of the dwarfs from the fitted plane. The system is viewed from infinity and rotated by angle $\phi=158.0^{\circ}$ so that the VPOS is viewed edge on.}
\label{vpos}
\end{center}
\end{figure}

At the current time, the positions of every Milky Way dwarf (both confirmed and unconfirmed) and the best-fitting plane are shown in Fig. \ref{vpos}. The hypothetical plane is fitted only to the 11 `classical' dwarfs or those dwarfs for which we have proper motion measurements. The data plotted in Fig. \ref{vpos} are rotated by an angle of $\phi=158.0^{\circ}$ about the $z$-axis so that the best-fitting plane is viewed edge-on. Using a clockwise rotation matrix, the rotated coordinates are found by $x_{\text{rot}}=x\cos{\phi}+y\sin{\phi}$ and $y_{\text{rot}}=-x\sin{\phi}+y\cos{\phi}$.

\begin{figure}
\begin{center}
\includegraphics[width=\columnwidth]{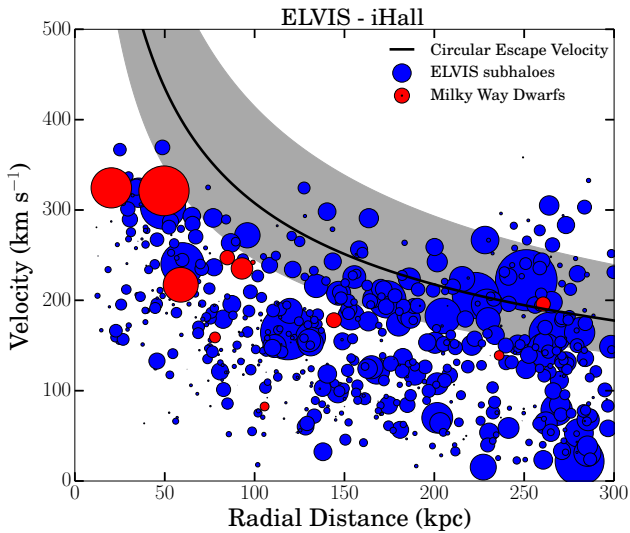}
\caption{The Milky Way classical dwarfs are shown in red, while the sub-haloes remaining after our velocity and distance cuts from the ELVIS iHall simulation are shown in blue. The size of a point is related to its mass and has been normalized by the LMC, the most massive Milky Way dwarf. The black line shows the circular escape speed for the Milky Way at $M_{\text{MW}}=1.1\times10^{12}\text{M}_{\odot}$, and the grey shaded region represents the circular escape speed for the Milky Way in the mass range $(0.7--2)\times10^{12}\text{M}_{\odot}$. }
\label{iHall_vr}
\end{center}
\end{figure}

The same plane-fitting analysis is also performed on the dark matter simulations described above. To make an accurate comparison to the Milky Way, we perform a series of cuts to the sub-haloes found at the end of each simulation and limit our analysis only to those sub-haloes capable of holding baryonic matter. We limit the sample by selecting sub-haloes that have $V_{\text{peak}} > 10$ km s$^{-1}$, $V_{\text{peak}}$ being the maximum circular velocity at the point when the sub-halo contained the most mass, and are within 300 kpc. This ensures that we consider all sub-haloes that have enough mass to be considered a dwarf and are within the virial radius of the Milky Way. The resulting total number of sub-haloes for each simulation is displayed in Table \ref{DM_sim_values}, and Fig. \ref{iHall_vr} visually shows the distribution of sub-haloes that survive our data cuts for the iHall simulation. From these remaining sub-haloes, 11 are chosen at random to represent the 11 classical dwarfs and a hypothetical plane is fitted to their positions. This process is then repeated $10^5$ times for each simulation.

When considering the VPOS, \citet{Kroupa:2005aa} looked at subsets of the dwarf population to see how plane-like the structure remained with fewer satellites. This was done by calculating the angle $|\cos(\omega)|$, which is calculated here by finding the angle between the normal of the best-fitting plane and the vector pointing to a sub-halo. It is a measurement of the `thinness' of a plane where a perfectly planar distribution would result in $|\cos(\omega)|=0$ for every object. They ultimately found that the plane-like structure was extremely unlikely to have formed from an isotropic parent distribution. This analysis was revisited by \citet{Zentner:2005aa}, who showed that the null hypothesis posited by \citet{Kroupa:2005aa} was incorrect. \citeauthor{Kroupa:2005aa} had used the $|\cos(\omega)|$ angle from $10^5$ satellites: a nearly uniform distribution and correct only in the limit of large sample size. \citet{Zentner:2005aa} corrected this by taking a small subset of an isotropic distribution and then comparing $|\cos(\omega)|$ to that of the observed VPOS. With the correct null hypothesis, it was found that the Kolmogorov--Smirnov (KS) probability of drawing the Milky Way satellites from the sample of CDM sub-haloes was $P_{\text{KS}}\simeq 0.15$. This was further improved by comparing to a more realistic triaxial halo where it was found that the KS probability was even higher. We repeat this analysis here using the two-sample KS test comparing the values of $|\cos(\omega)|$ between the observed dwarfs and CDM sub-haloes.

\subsection{Orbit calculations: the VPOS as a function of time}

We also perform orbit calculations using the observed proper motions of the classical satellites. We use the fourth-order Runge--Kutta orbit integrator code developed by \citet{Chang:2011aa} to calculate the orbital distribution over time. We employ a static, spherical, \citet{Hernquist:1990aa} potential that has the same mass and inner density slope within $r_{200}$ as an equivalent NFW \citep{Navarro:1997aa} profile. The host halo has a mass of $M=1.1\times10^{12} \text{M}_{\odot}$ \citep{Watkins:2010aa, Deason:2012aa, Wang:2012aa} and a concentration parameter of $c=9.39$. These orbit calculations match well with the dark matter simulations described above, which use NFW halo solutions for their host galaxies. The ELVIS haloes have concentration values of $c_{\text{iHall}}=5.8$ and $c_{\text{iScylla}}=9.5$ \citep{Garrison-Kimmel:2014aa}. 

The potential is described by

\begin{equation}
\Phi(r)=-\frac{GM_\text{T}}{r+a},
\label{potential}
\end{equation}

where $a$ is the scale-length of the Hernquist profile and $M_\text{T}$ is the normalization to the potential or the total mass. Dynamical friction is also modelled using the Chandrasekhar formula \citep{Besla:2007aa, Chang:2011aa} and the equation of motion has the form

\begin{equation}
\ddot{\bm{r}}=\frac{\partial}{\partial r}\Phi_{\text{MW}}(|\bm{r}|)+\bm{F}_{\text{DF}}/M_{\text{sat}},
\end{equation}

where $M_\text{sat}$ is the satellite mass, $\Phi_{\text{MW}}(|\bm{r}|)$ is the potential corresponding to equation \ref{potential} and $\bm{F}_{\text{DF}}$ is the dynamical friction term. The dynamical friction term is given by:

\begin{equation}
\bm{F}_{\text{DF}}=-\frac{4\pi G^2M^2_{\text{sat}}\ln(\Lambda)\rho(r)}{v^2}\left[\erf(X)-\frac{2X}{\sqrt{\pi}}\exp(-X^2)\right]\frac{\bm{v}}{v}.
\end{equation}

Here, $\rho(r)$ is the density of the dark matter halo at a galactocentric distance, $r$, of a satellite of mass $M_\text{sat}$ travelling with velocity $v$; $X=v/\sqrt{2\sigma^2}$, where $\sigma$ is the 1D velocity dispersion of the dark matter halo, which is adopted from the analytic approximation of \citet{Zentner:2003aa}. The Coulomb logarithm is taken to be $\Lambda = r/(1.6k)$, where $k$ is the softening length if the satellite is modelled with a Plummer profile.

\subsubsection{Masses and proper motions} 
 For Carina, Draco, Fornax, Leo I, Leo II, Sculptor, Sextans, and Ursa Minor, we consider the total dynamical mass out to the maximum radius of the velocity dispersion data (an estimate of the total mass) from \citet{Walker:2009aa}. The Sagittarius dwarf galaxy is in the process of being tidally disrupted and equilibrium mass estimators are particularly ill-suited for this system. Cited progenitor masses have varied over two orders of magnitude for the Sagittarius dwarf \citep{Johnston:1999aa, Law:2010aa, Purcell:2011aa}. \citet{Chakrabarti:2014aa} addressed this issue and showed that one may derive joint constraints on the progenitor masses of tidally disrupting satellites by employing the current position and velocity of the Sagittarius dwarf and its maximum radial excursion (estimated from observed tidal debris). The progenitor mass estimate we adopt from \citet{Chakrabarti:2014aa} for the Sagittarius dwarf is $1 \times 10^{10} \text{M}_{\odot}$. For the LMC and SMC, we adopt mass estimates from \citet{van-der-Marel:2014aa}, which are $3 \times 10^{10}$ and $3 \times 10^{9} \text{M}_{\odot}$ respectively. 

Proper motion data for Ursa Minor, Sculptor, Sextans, Carina, Fornax, and Leo I were gathered from \citet{Sohn:2013aa}. The proper motion measurements for Draco have been vastly improved since prior studies, and we have adopted the values reported in the recent work by \citet{Casetti-Dinescu:2016aa}. This is also true of the motion of Leo II, which was recently reported by \citet{Piatek:2016aa} to have a proper motion measurement of less than half of its previously reported value. Proper motion values from the LMC and SMC were taken from \citet{Kallivayalil:2013aa}. Finally, for Sagittarius, we adopt the values reported by \citet{Massari:2013aa}, which contain the best estimate of the mean centre-of-mass motion. Their measurements correct for the fact that Sagittarius is large enough on the sky that there are perspective differences for the different fields that are measured and consider the most accurate proper motion measurements of other works as well. The errors of these measurements were obtained by sampling the error in the proper motion measurements directly in $10^4$ Monte Carlo realizations, assuming a Gaussian distribution, and then converting into $V_x$, $V_y$, and $V_z$ (Sohn, private communication).

\section{Members of the VPOS}

\begin{figure}
\includegraphics[width=\columnwidth]{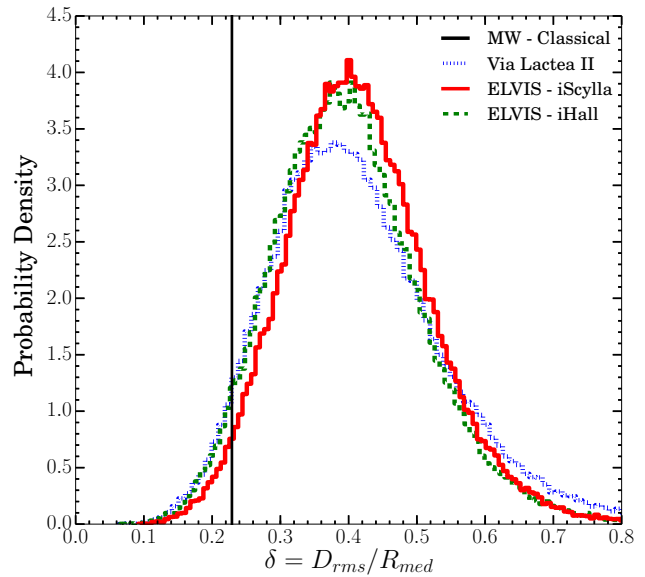}
\caption{The probability density that a plane fitted to a random distribution of 11 haloes will have a certain thickness for the three dark matter simulations considered in this paper. Thinner distributions will appear to the left-hand side. The Milky Way distribution is shown as the vertical black line and appears to be thinner than most random distributions from dark matter simulations.}
\label{delta_plot}
\end{figure}

Performing the plane-fitting analysis to all the classical dwarfs results in $D_{\text{rms}}=21.3$ kpc or $\delta_{\text{Classical}} = 0.23$. For $10^5$ samples of plane fitting in dark matter simulations, the VLII simulation has $\delta_{\text{VLII}} = 0.41\pm0.13$, iHall has $\delta_{\text{iHall}} = 0.39\pm0.11$, and iScylla has $\delta_{\text{iScylla}} = 0.41\pm0.11$. These results are graphically shown in Fig. \ref{delta_plot}. Thus, the VPOS appears more constrained than the average cosmological simulation but is still well within $2\sigma$ of the mean of all the distributions. 

The alignment of the VPOS is also more polar than most cosmological simulations predict. We define the angle between fitted planes and the host halo disc as $\theta$ and find that $\theta=77.3^{\circ}$ for the VPOS in agreement with other work \citep[e.g.][]{Shao:2016aa}. Planar structures found in simulations tend to lie at shallow angles compared to their host halo, often at angles $\theta<45^{\circ}$ \citep{Zentner:2005aa}. Since the simulations used in this analysis consist of only dark matter, a similar comparison cannot be made here. 

The KS test reveals the likelihood that two distributions are identical. Here we compare the values of $|\cos(\omega)|$ obtained from the Milky Way distribution with the values of $|\cos(\omega)|$ from each of the $10^5$ distributions obtained from simulations and report the mean value. For VLII, we obtain a value of $P_{\text{KS,VLII}} = 0.50$, iHall has $P_{\text{KS,iHall}} = 0.59$, and iScylla has $P_{\text{KS,iScylla}} = 0.59$. This shows that it is very likely that the two distributions come from the same underlying distribution. The above results have been summarized in Table \ref{results}. The probability, $P$, of drawing a thinner distribution, $z$, from a simulation compared to the observed one is found by computing the total number of thinner distributions over the total. Fig. \ref{delta_plot} shows that the Milky Way dwarfs are less planar than 4.2 per cent of CDM distributions, or, in other words, there is about a 1 in 24 chance of drawing a more planar distribution from a CDM simulation.

Previous work by \citet{Libeskind:2014aa} has shown that the most massive sub-haloes have a much higher degree of anisotropy than smaller sub-haloes due to a preferential infall with respect to large-scale structure. When we limited our samples to include only the 50 most massive dwarfs, we also find that our random distributions become more planar. For the classical dwarfs, this translates into a 1 in 10 chance of drawing a more planar distribution, more than doubling the probability when compared to the full sample. By limiting the distribution to contain only massive dwarfs, we find a leftward shift in the distributions of Fig. \ref{delta_plot}. For the analysis in this paper, we choose to use the full sample of sub-haloes described in Table \ref{DM_sim_values}.

\begin{figure}
\begin{center}
\includegraphics[width=\columnwidth]{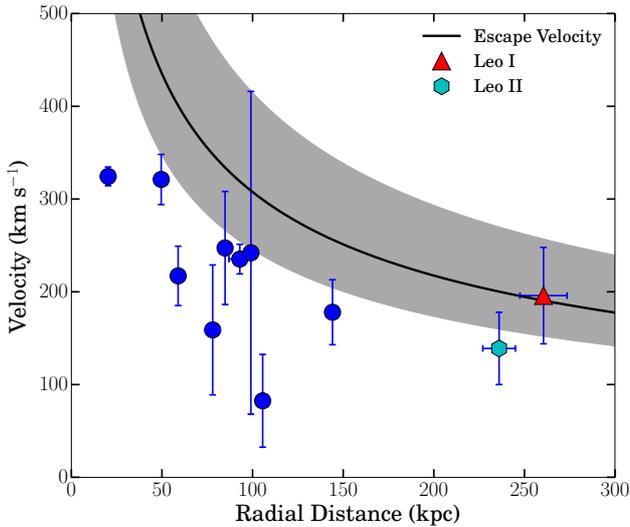}
\caption{The measured velocities of the classical dwarf galaxies are shown as a function of their radial location. The black line shows the circular escape speed for the Milky Way at $M_{\text{MW}}=1.1\times10^{12}\text{M}_{\odot}$, and the grey shaded region represents the circular escape speed for the Milky Way in the mass range of $(0.7--2)\times10^{12} \text{M}_{\odot}$. Leo I and Leo II are travelling close to or in excess of the escape velocity and lie at a much greater distance than the other classical dwarfs.}
\label{f:dwarf_velocities}
\end{center}
\end{figure}

 While it does appear that the VPOS is perhaps an uncommon structure, it does not appear to be unique within our results. To understand what drives the uncommon nature of the structure, we look at the population of classical dwarfs and consider a few individual members. Subsets of the VPOS have been considered by others; for example, \citet{Pawlowski:2013aa} and \citet{Kroupa:2005aa} considered subsets of the classical dwarf population based on orbital pole calculations. Here we will look at the positions and velocities of the dwarfs to analyse subsets. Fig. \ref{f:dwarf_velocities} shows the velocities of each dwarf and their radial distance. 

It is clear from Fig. \ref{f:dwarf_velocities} that all the classical dwarfs, except Leo I and Leo II, lie close to the Milky Way centre and are travelling within the escape velocity. The two dwarfs lie at distances greater than 200 kpc and therefore may drive a tight plane solution. Therefore, Leo I and Leo II appear to be outliers in both position and velocity and deserve closer inspections. Also, depending on the choice of mass for the Milky Way, more dwarfs may loosely bound.

\subsection{Leo I}
A number of authors have argued that Leo I's extreme kinematic properties indicate that it is not a bound satellite of the Milky Way \citep{Sales:2007aa, Mateo:2008aa, Rocha:2012aa, Pawlowski:2013aa, Sohn:2013aa}. \citet{Sohn:2013aa} performed a proper motion study of Leo I based on two epochs of \textit{Hubble} ACS/WFC images separated by $\sim$5 yr time. In their study, they examined the proper motion error space, the star formation history, and the interaction history of Leo I. \citeauthor{Sohn:2013aa}'s analysis also showed that Leo I is most likely on first infall, having experienced only one pericentre passage within a Hubble time and appears to be on a parabolic or a nearly bound orbit. Orbit integrations of Leo I match up extremely well with epochs of star formation. A burst of star formation $\sim$2 Gyr ago corresponds with the moment that Leo I entered the virial radius of the Milky Way. This is then followed by a quenching of star formation $\sim$1 Gyr ago matching well with a pericentre passage. At the point of pericentre, Leo I would have experienced significant ram pressure stripping, removing the leftover gas from the halo. \citet{Pawlowski:2013aa} further showed through orbital pole analysis that Leo I is not aligned with the other satellites of the VPOS and therefore should not be included in analysis.

\subsection{Leo II}

 Until recently, the proper motion measurements of Leo II were poorly constrained. It was believed that Leo II had a large tangential velocity \citep[$265\pm129$ km s$^{-1}$;][]{Lepine:2011aa}, which led to an overall velocity measurement of $266\pm129$ km s$^{-1}$. This measurement has been highly refined by \citet{Piatek:2016aa}, resulting in a new tangential velocity of $127\pm42$ km s$^{-1}$ and a total velocity of $129\pm39$ km s$^{-1}$, less than half of the previous measurement but reasonably within the large error range of previous studies. Due to the previous large errors in velocity, \citet{Rocha:2012aa} ignored the velocity measurements and instead looked at the star formation history of Leo II and found that star formation occurred in Leo II up until $\sim$2 Gyr ago. This led the authors to the conclusion that infall occurred between $\sim$2 and 6 Gyr ago, indicating a fairly recent merger.

With this new velocity information, we investigate the inclusion of Leo II in a plane-fitting analysis by calculating its past orbital history. The Hernquist model was used in our calculations as described above; however, it was found that when a purely NFW profile or singular isothermal sphere (SIS) profile was used, it had little impact on results. This is due to the large distance at which Leo II interacts with the Milky Way; at such large distances, the different inner slopes of the profiles do not affect the motion of the satellite which mostly sees the Milky Way as a point source. The inclusion of dynamical friction in our model also showed very little effect due to the large distance of Leo II and its small mass. Furthermore, for different Milky Way masses, the results remained nearly unchanged.

\begin{figure}
\includegraphics[width=\columnwidth]{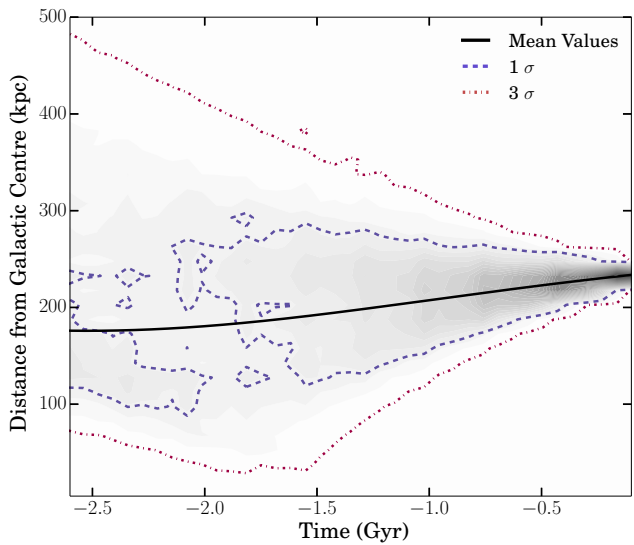}
\caption{Shown are the results of $10^3$ realizations of the orbital history of Leo II sampling the proper motions and uncertainties thereof, from $t=-2.6$ Gyr to the present day ($t=0$). The grey-scale contours show the relative probability density of an orbit being in a certain location at a given time, and the lines represent the $1\sigma$ (blue, dashed) and $3\sigma$ (red, dash--dotted) contours within which the orbit realizations are contained. The black line corresponds to the realization using the mean measured values.}
\label{LeoII_hist}
\end{figure}

 Using the new velocity data and errors from \citet{Piatek:2016aa} for Leo II (Table \ref{dwarf_data}), we sampled the $\pm3\sigma$ velocity error space $10^3$ times and integrated each realization backward in time. Only the velocity error space was investigated since the position measurement of Leo II is well known. Fig. \ref{LeoII_hist} shows the galactocentric location of Leo II as a function of time, the $1\sigma$ and $3\sigma$ probability density contours, and the relative probability density contours. Due to errors in velocity, the results diverge at late times, but, in general, most realizations show that Leo II has spent most of its time at distances similar to its current position. At most, Leo II appears to have made only one pericentre passage within the last 2.5 Gyr. Therefore, it seems likely that Leo II is on first infall or has a similar orbital history to Leo I.

\subsection{Other dwarfs}

From Fig. \ref{f:dwarf_velocities}, it appears that there are five other dwarfs that may be close to the escape velocity if the Milky Way is very light ($M_{\text{MW}}<1\times10^{12} \text{M}_{\odot}$). These galaxies include LMC, Sculptor, Draco, Sextans, and Fornax. 

From an analysis of radial position, velocity information, star formation history, and comparisons to VLII data, \citet{Rocha:2012aa} were able to estimate the infall times for the classical dwarfs. Based on radial positions and velocities alone, Draco and Sextans likely fell on to the Milky Way more than 8 Gyr ago. From proper motion measurements, the small tangential velocity of Fornax and Sculptor disfavours a recent infall scenario and instead indicates an infall time of more than 5 and 8 Gyr ago, respectively. Furthermore, the presence of old stellar populations with no evidence of stars younger than $\sim$10 Gyr indicates that Draco and Sculptor have been long-time satellites of the Milky Way having experienced ram pressure and tidal stripping long ago ($t_{\text{infall}}>8$ Gyr). These results indicate that Draco, Sextans, Fornax, and Sculptor are old mergers that may be part of a long lived structure if it exists.

When examining the proper motions and local environment of the LMC and SMC, studies have found that they are most likely a binary or active merger that is falling on to the Milky Way for the first time \citep{Besla:2007aa, Besla:2010aa, Kallivayalil:2013aa}. \citet{Rocha:2012aa} also found that the large 3D velocity and active star formation seen in the LMC demands a recent merger history of less than 4 Gyr ago. This indicates that the LMC and SMC may not be part of a long-lived structure.

\subsection{Plane fitting revisited}
When we perform the plane-fitting analysis again without including Leo I and Leo II, we find only a slightly different distribution (Table \ref{results}). The thickness of the plane decreases to $D_{\text{rms}}=20.7$ kpc; however, it is now more compact, and therefore the normalized thickness grows to $\delta = 0.24$. This corresponds to a probability of randomly drawing a thinner distribution to about 1 in 19. Another interesting point to note is that the angle between the fitted plane and the Milky Way disc becomes steeper and is nearly perpendicular. Therefore, it appears that Leo I's and Leo II's distant radial locations do not influence a specific polar fit to the VPOS as a similar one is seen even without their inclusion. 

If instead we look at only the dwarfs that have aligned angular momentum vectors as pointed out by \citet{Kroupa:2005aa} and \citet{Pawlowski:2013aa}, we find almost the same solution as in the case of the classical dwarfs. All of the above distributions are pictured graphically in the lefthand column of Fig. \ref{distributions}. 

The probabilities that are seen in this study are similar to $\sim$2$\sigma$ of a normal distribution, which matches fairly well with the distributions in Fig. \ref{delta_plot}. This indicates that planar distributions do not appear to be rare in $\Lambda$CDM simulations. \citet{Pawlowski:2014aa} have also performed an analysis of ELVIS data, but found that the structures in that data set did not match the VPOS. However, their analysis did not consider the compact distribution of the Milky Way sample. Their sub-halo selection also differed from this one as they chose only the largest 11 satellites remaining after obscuring a region to correspond to a galactic disc, while the analysis in this paper is left open to a random sampling of the sub-haloes remaining after imposing a cut on $V_{\text{peak}}$ and radial distance. Furthermore, the ELVIS simulations used by \citet{Pawlowski:2014aa} were not of a high enough resolution to resolve sub-haloes that are comparable to the smaller classical dwarfs such as Leo II.

\begin{table*}
\caption{Plane-fitting results. Column 1 gives the distribution on which the analysis is done. Column 2 gives the $D_{\text{rms}}$ thickness of a theoretical plane fit to sub-haloes. For the Milky Way, this is done to the classical or remaining dwarfs; for dark matter simulations, this is done to a random drawing of 11 sub-haloes, as described in the text. Column 3 gives a measure of the thickness as normalized by the median distance of the distribution, as defined by equation \ref{delta}. Column 4 shows the probability, $P$, of randomly drawing a distribution, $z$, from a simulation that is thinner than the observed distribution. Column 5 gives the KS probability that the Milky Way sample is from the same distribution as dark matter simulations. Column 6 is the angle between the theoretical plane fit to the Milky Way dwarfs and the Milky Way stellar disc. }
\begin{center}
\begin{tabular}{cccccc}
\hline\hline
 Distribution& $D_{\text{rms}}$ (kpc) & $\delta$ 	& $P(z<\delta$)	&$P_{\text{KS}}$ & $\theta(^{\circ})$	\\ \hline
VLII		& 37$\pm$13	& 0.41$\pm$0.13	& \textendash	& 0.50		& 	\textendash	\\
iHall		& 61$\pm$18	& 0.39$\pm$0.11	& \textendash	& 0.59		&	\textendash	\\
iScylla	& 64$\pm$17	& 0.41$\pm$0.11	& \textendash	& 0.59		&	\textendash	\\ \hline
 Classical	& 21.3		& 0.23			& 1 in 24 		& \textendash 	&	 77.3			\\
 No Leo I/II	& 20.7		& 0.24			& 1 in 19		& \textendash 	&	 85.3			\\
 Aligned (PK13)	& 22.6 	& 0.23 			& 1 in 24		& \textendash 	& 79.8\\
 Trusted PMs & 20.1 		& 0.21 			& 1 in 40		& \textendash 	& 79.2\\ \hline
\end{tabular}
\end{center}
\label{results}
\end{table*}

It is also worth while to discuss the other members of the VPOS as much work has been done investigating their inclusion. As mentioned earlier, the proper motions and local environment of the LMC and SMC indicate that they are most likely a binary or active merger that is falling on to the Milky Way for the first time. Other dwarfs appear to have orbits that carry them away from the VPOS, such as Ursa Minor \citep{Piatek:2005aa} and Fornax \citep{Piatek:2007aa}. Orbital pole analysis has also shown that Sagittarius is on a polar orbit but lays at an approximate right angle to both the VPOS and the Milky Way disc \citep{Palma:2002aa}, which may have been caused by it being scattered into it's current position by an encounter with the LMC/SMC \citep{Zhao:1998aa}.

Without including the above-mentioned dwarfs, there remains only four dwarfs not affected by these results: Sculptor, Draco, Sextans, and Carina. Since the minimum number of objects needed to perform a KS test is six and fitting a plane to four objects leads to little insight, we do not perform further analysis on the reduced population.

\section{Stability of the VPOS}
 
\begin{figure*}
\includegraphics[width=0.98\textwidth]{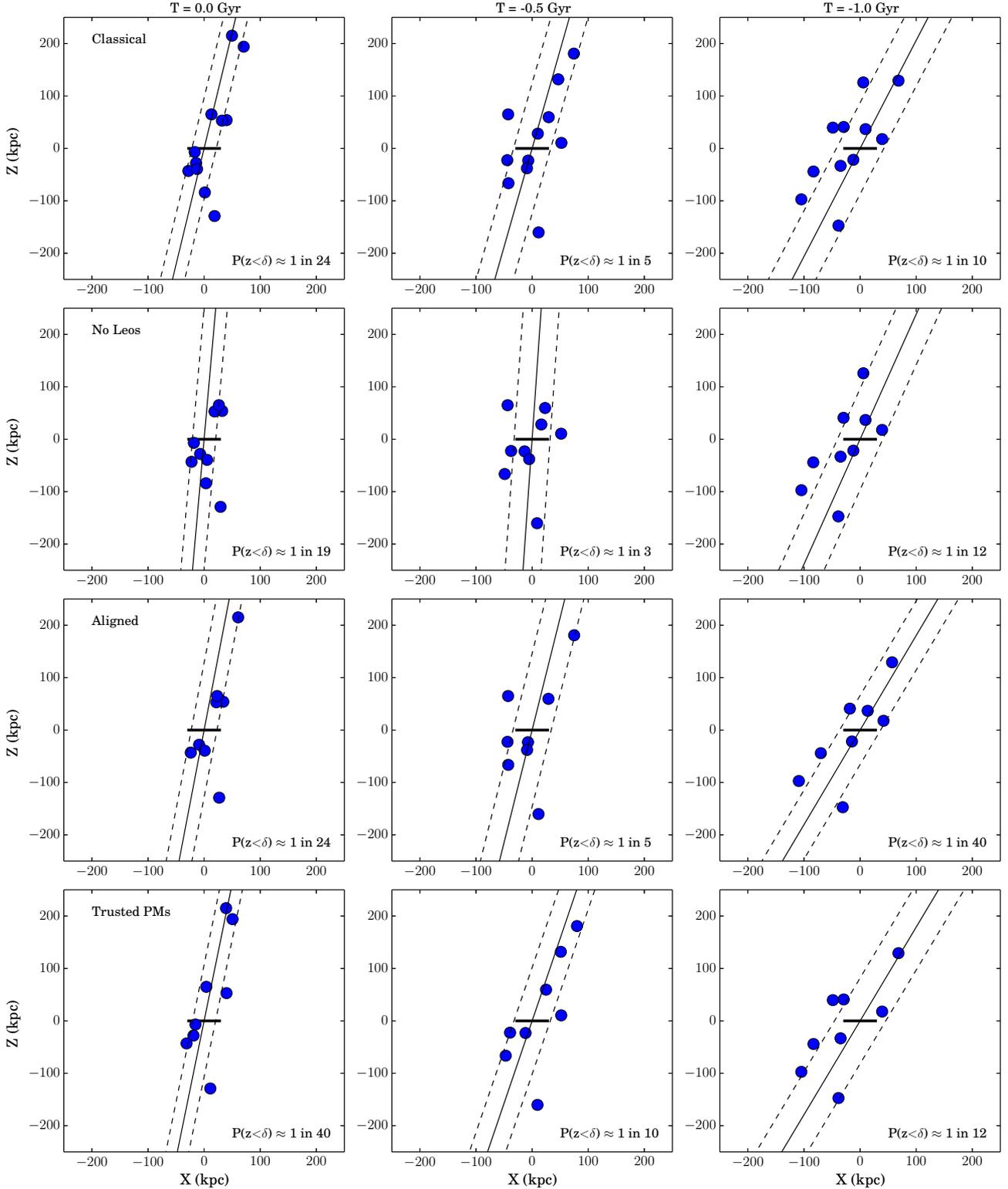}
\caption{ Each row represents a different subset of the Milky Way classical dwarfs integrated backward in time. Due to errors in the proper motion measurements, these integrations cannot be trusted beyond 0.5 Gyr (see Section 4.1). First row: all 11 classical dwarfs. Second row: the classical dwarfs minus Leo I and Leo II. Third row: only the classical dwarfs that have aligned angular momentum vectors as defined in \citet{Pawlowski:2013aa}. Fourth row: only the classical dwarfs that have proper motion measurements accurate enough to be trusted in orbit integrations beyond 0.5 Gyr. This is the only distribution that does not show a dramatic increase in significance at $t=-1$ Gyr. In all panels: blue dots represent Milky Way dwarfs, the thick black line represents the Milky Way stellar disc, the thin black line represents the hypothetical best-fitting plane viewed edge on, the dashed lines represent the rms thickness of the best-fitting plane, and the probability of drawing a more planar distribution from a simulation is shown (see Section 3). The simulation time is shown at the top of each column. }
\label{distributions}
\end{figure*}

Another important question to address is whether or not the VPOS is a long-lived structure or a temporary alignment. For the 11 classical dwarfs, proper motion measurements have been made and are listed in Table \ref{dwarf_data}. Using the test particle code developed by \citet{Chang:2011aa} described above, we have simulated the orbits of the classical dwarfs backward in time (Fig. \ref{distributions}, top row). Also shown in Fig. \ref{distributions} are the orbit integrations when considering the classical dwarfs minus Leo I and Leo II (second row) and the distribution of dwarfs that have aligned orbital poles (third row).

The plane-fitting solution is essentially the same at the current time in all cases, with a nearly polar fit to the distribution and similar probabilities for drawing a more planar distribution from a simulation (Fig. \ref{distributions}, left-hand column). For every case, within 0.5 Gyr (half a dynamical time), the distribution becomes wider, and the likelihood of drawing a more planar distribution increases by a factor of $\sim$5. At 1 Gyr, all the distributions appear to grow slightly thinner and tend towards a more coplanar solution. 

 \subsection{The influence of observational errors in orbit calculations}
Since the errors in the proper motion measurements are quite large for some of the dwarfs we investigate further to understand how much observational errors influence our results. To test this, we set up a ring of 11 evenly spaced particles at radius $r=150$ kpc with a circular velocity. The simulation was then run for 10 Gyr to ensure that stability was maintained. The dynamical time for a ring of radius 150 kpc is:

\begin{equation}
\tau_{\text{dyn}} \sim \tau_{\text{cross}} \sim \frac{R}{v} \sim 0.9 \text{ Gyr}
\end{equation}

We begin with a ring of particles and introduce simulated observational errors which inevitably causes our perfect ring of particles to disperse. As a result, the ring will grow in width and the amount by which the distribution is constrained to the shape of a ring is defined by Equation \ref{delta}. Eventually, at very late times and with large errors the particles will disperse to the extent that their distribution is better described as a disc rather than a coherent ring. The moment when this transformation takes place is best described as when the $3\sigma$ thickness of the ring is comparable to the median radial extent of the particles. We therefore define the criteria that $\delta < 0.33$ describes a well-constrained ring. 

Introduction of errors is done by percentages; we assume that there is a certain amount of error in each component of radius and velocity. We then randomly sample the $\pm3\sigma$ error space associated with a percentage of the true value for each component and repeat the simulation 100 times for each percentage, the results are given in Table \ref{error_prop}. The \textit{HST} proper motion measurements have larger errors in velocity than in the positions of the dwarfs. Therefore, we use the average error in position and only vary the percentage error in velocity. To examine the impact of the measured position errors, we simulated the worst-case scenario based on Draco, which has a radial distance of $93\pm4$ kpc corresponding to an error of 4.3 per cent. When this percentage was used for each component of position with no velocity errors, almost no impact on the ring structure was found. The median error of each position component for the classical dwarfs is $\sim$3.5 per cent; we therefore used this for the position errors for all simulations. 

It was found that even with 100 per cent error in each component of velocity, the ring maintained coherence past $t=-0.5$ Gyr. For velocity errors up to 50 per cent, the ring maintained coherence past $t = -0.75$ Gyr, and for errors up to 30 per cent, the ring maintained coherence longer than a dynamical time. Looking at the velocity measurements of the classical dwarfs, we see that 8 of the 11 have velocity errors that are under 100 per cent in each component. Ursa Minor, Sculptor, and Carina fail this criteria. Interestingly, Sextans has the largest error in velocity measurements but passes our criteria for being trustworthy up to 0.5 Gyr. Only three dwarfs pass the criteria for integrations lasting up to 1 Gyr: Sagittarius, LMC, and Draco. All other dwarfs have large uncertainties in general or have at least one small velocity component that is very uncertain. 

These results indicate that 8 of the 11 classical dwarfs have errors that can be trusted in integrations beyond 0.5 Gyr: Sagittarius, LMC, SMC, Draco, Sextans, Fornax, Leo II, and Leo I. To reduce the amount of error entering our orbit calculations and to test whether the plane loses significance due to a few high-error dwarfs, we restrict ourselves to the dwarfs mentioned above and integrate the orbits again to see if coherence is maintained (Fig. \ref{distributions}, bottom row). We find that they too more or less follow the same trends as the other distributions. Interestingly, however, at $t=-1$ Gyr, we find that the significance of the plane remains the same rather than increasing as the other populations do. 

These results also indicate that our integrations cannot be fully trusted up to 1 Gyr, especially those that contain the high-error dwarfs, Ursa Minor, Sculptor, and Carina, but can be trusted up to 0.5 Gyr. Therefore, the trends that are seen at $t=-0.5$ Gyr are indicative of the true solution. 
 
 \begin{table}
\caption{The first column indicates the amount of error given to each component of velocity of our test particles, while position errors were held at the median position component error of the Milky Way dwarfs: 3.5 per cent. Columns 2--4 give the normalized thickness of the ring at $t=-0.5, -0.75,$ and $-1$ Gyr respectively. Those values shown in bold fail our criteria for coherence.}
\begin{center}
\begin{tabular}{cccc} \hline\hline
\makecell{Per cent error in each \\ velocity component}	& $\delta_{t=-0.5\text{Gyr}}$& $\delta_{t=-0.75\text{Gyr}}$ 	&$\delta_{t=-1\text{Gyr}}$ \\ \hline
10	& 0.05		& 0.08		& 0.11	\\
20	& 0.08		& 0.14		& 0.20	\\
30	& 0.11		& 0.19		& 0.30 	\\
40	& 0.14		& 0.26		& \textbf{0.38}	\\
50	& 0.18		& 0.31		& \textbf{0.42}	 \\
60	& 0.20		& \textbf{0.35}	& \textbf{0.52}	 \\ 
70	& 0.23		& \textbf{0.40}	& \textbf{0.56}	 \\ 
80	& 0.26		& \textbf{0.43}	& \textbf{0.59}	\\ 
 90	& 0.28		& \textbf{0.47}	& \textbf{0.64}	\\ 
 100	& 0.31		& \textbf{0.50}	& \textbf{0.69}	\\ \hline
\end{tabular}
\end{center}
\label{error_prop}
\end{table}

\subsection{Limitations}

There are many factors that may influence the outcome of our orbit integrations. First, the shape of the Milky Way halo is currently very uncertain. It is almost certainly non-spherical, but the extent and the orientation of the halo has varied significantly in studies \citep[e.g.][]{Law:2010aa, Loebman:2012aa, Debattista:2013aa, Deg:2014aa, Bovy:2016aa}. There are also observed differences between the shapes of haloes in simulations compared to observations since simulations mainly find triaxial haloes and observations mainly find spherical haloes \citep{Debattista:2008aa}. It appears that the presence of baryons condensing within haloes works to make haloes rounder, at least in their central regions \citep{Debattista:2013aa}. It may be the case that the haloes have spherical cores that become triaxial at large distances, in which case, a non-spherical halo may work to shepherd sub-haloes into a preferred axis. \citet{Shao:2016aa} studied how likely it was to have satellite systems nearly perpendicular to the disc of their central galaxies, and found that it occurred $\sim$20 per cent of the time in simulations. Here we have assumed spherical haloes; however, the plethora of proper motion data that \textit{Gaia} will deliver should serve to constrain the true shape of the Milky Way potential in the near future \citep{Price-Whelan:2013aa, Price-Whelan:2014aa, Bovy:2016aa}.

Another factor that influences our simulations is our choice of Milky Way halo mass, which is also very poorly constrained. Through a multitude of methods, the Milky Way mass has been estimated to be $\sim(0.7-2)\times 10^{12}\text{M}_{\odot}$ \citep{Watkins:2010aa, Deason:2012aa, Wang:2012aa, Boylan-Kolchin:2013aa}, although measurements can vary quite dramatically \citep[see fig. 1 of][]{Wang:2015aa}. Many of these estimates use measurements of the mass from the inner 10--50 kpc and extrapolate to the full halo and are therefore highly effected by the choice of the Milky Way halo shape. To investigate the impact of our choice of halo mass, we varied the mass in the range of $\sim(0.7-2)\times 10^{12}\text{M}_{\odot}$. At $t=-0.5$ Gyr, almost no effect between simulations was seen; however, by $t=-1$ Gyr, the effect was visible. For lower halo mass simulations, the distribution was significantly more dispersed than higher halo mass simulations. With a lower Milky Way mass, satellites experience longer orbits as more dwarfs are travelling closer to the escape velocity. Therefore, it seems that between the limitations imposed by observational errors in proper motion measurements and the uncertainty in halo mass, orbit integrations can be trusted only up to $\sim$500 Myr.

\section{Discussion}

 We have compared the Milky Way distribution of massive dwarf galaxies to distributions of sub-haloes in CDM simulations. It is interesting to note that a relatively few sub-haloes in simulations are travelling in excess of the escape velocity, while it appears that perhaps two dwarfs of the Milky Way do. VLII, for example, has almost none ($\sim$1 per cent) and Fig. \ref{iHall_vr} shows that the ELVIS simulations form relatively few; furthermore, recent work by \citet{Boylan-Kolchin:2013aa} found almost none in the simulations that they analysed. It is important to consider if high-velocity dwarfs near the escape velocity of their hosts contribute to plane-like structures due to their large radial extents. When the high-velocity, distant dwarfs of the Milky Way were removed (Leo I and Leo II), we found almost no difference in our plane-fitting solution. Also, when we limit our analysis to include only random selections that include at least one high-velocity dwarf, we find that structures become only marginally more planar. Although, this may be connected to another underlying discrepancy between simulations and observations: Milky Way dwarfs are more centrally located. Simulations tend to have most sub-haloes at large distances, while the Milky Way dwarfs are tightly clustered near the centre, nine within 150 kpc \citep{Kravtsov:2004aa}. This effect can be seen in Fig. \ref{iHall_vr} where the largest concentration of sub-haloes is at a large radial distance, this is especially evident with massive sub-haloes. In this study, we have limited our halo selection to choose only those haloes that have $V_\text{peak}>10$ km s$^{-1}$, which helps to remove a radial bias from our sample that would be present if we limited our sample by $V_\text{max}$, the maximum circular velocity of a sub-halo at the current time. However, this still leads to a slightly radially extended solution because of resolution issues near the centres of massive haloes. Although, \citet{Kravtsov:2004aa} were able to match the Milky Way distribution well with their model, which found that only the largest sub-haloes were able to accrete enough matter to form successful galaxies before being tidally stripped to the smaller masses we observe today.

It must be realized that there exists some observation bias to observe only those dwarfs above and below the plane of the Milky Way \citep{Mateo:1998aa}, although these areas are now starting to be explored \citep{Chakrabarti:2015aa, Chakrabarti:2016aa}. The current dearth of Milky Way satellites close to the plane \citep{Mateo:1998aa} and that lack of observational surveys probing Galactic sub-structure at low latitudes may mean that we have missed some massive Milky Way dwarfs close to the Galactic plane \citep{Chakrabarti:2009aa, Chakrabarti:2011aa}. Furthermore, there is evidence that the Milky Way sample is far from complete. \citet{Yniguez:2014aa} looked at the distribution of satellites around the Milky Way and M31, and found that within 100 kpc, the two distributions were remarkably similar but differed greatly at larger radii, indicating that our census of the bright Milky Way dwarfs is significantly incomplete beyond $\sim$100 kpc. \citet{Willman:2004aa} also argue that the Milky Way sample is incomplete due to obscuration and insufficient survey depth; therefore, the true number of dwarfs could be as much as three times larger than the current count. Recently, \citet{Homma:2016aa} reported the detection of an ultrafaint dwarf galaxy at a heliocentric distance of 87 kpc with the Subaru Hyper-Suprime Camera. They also point out that at the magnitude it was detected, SDSS has a completeness depth of only 28 kpc, meaning that even for dedicated surveys, dwarf galaxy detection is still incredibly difficult and far from complete. 

However, if deeper surveys do not uncover a similarly large number of bright satellites at large Galactocentric distances as we see in cosmological simulations, then we have to conclude that this is a real discrepancy. \citet{Kroupa:2010aa} have argued that this discrepancy suggests that a new solution using modified Newtonian dynamics (MOND) is needed on galactic scales to explain the distribution and number of satellites. Another solution is that torques from the baryonic components cause the satellites to lose angular momentum and spiral inwards. The majority of hydrodynamical, cosmological simulations do not produce sufficiently extended H\textsc{I} discs as in the observed Local Volume, so the lack of extended baryonic components in simulations (or the so-called angular momentum problem) may well be responsible for this problem \citep{Guedes:2011aa}. If this is the cause, then one would expect to see a similarly compact structure in high-resolution hydrodynamical cosmological simulations that \textit{do} have extended baryonic components, and it may be the case that spiral galaxies with extended H\textsc{I} discs have more compact satellite distributions. Therefore, an interesting simulation to consider is Illustris \citep{Genel:2014aa}, which includes both baryonic and dark matter. However, the resolution of Illustris allows only the detection of dwarfs down to a mass of $\sim 2\times10^7 \text{M}_{\odot}$ \citep{Haider:2016aa}. At this resolution, only galaxies larger than Ursa Minor would be detected. Simulations do not yet realistically represent physical processes that could make dwarfs spiral in more quickly or completely disrupt faster than expected. 

Planes of satellites similar to the observed Milky Way and M31 planes have been studied in detail in dark matter-only simulations by \citet{Buck:2015aa, Buck:2016aa}. Similar to this study, they found that these planes are not uncommon but are unstable and quickly disrupt. However, it has also been argued that the distributions of satellites found between dark matter-only and baryonic+dark matter simulations are inherently different \citep{Ahmed:2016aa}. Although the results in this paper contest with those found by \citet{Pawlowski:2014aa}, it is agreed that there is a need for high-resolution simulations that include both dark and baryonic matter.

The eight classical dwarfs that were found to have reliable proper motion measurements were shown to lose significance within 0.5 Gyr, which is within the time that their orbit integrations were found to be trustworthy. Beyond that, those dwarfs continued to show a low significance even up to 1 Gyr unlike the other distributions that were studied (Fig. \ref{distributions}). In order to more definitively say that the structure is not coherent, more accurate proper motions of all the dwarfs are needed, especially those dwarfs that have poorly constrained velocity components such as Ursa Minor, Sculptor, and Carina. However, in order to analyse the orbits of all the dwarfs beyond 0.5 Gyr, accurate proper motions of SMC, Sextans, Fornax, Leo II, and Leo I will also have to be achieved. All of these dwarfs have at least one velocity component that has an uncertainty nearly equal to or greater than it's magnitude. It appears that directionality is very important since the direction of a velocity component with a large uncertainty can flip, which will lead to very different solutions. 

Truly understanding the orbital history of a dwarf like Leo II is especially complicated. Because of its great distance from the Milky Way, large errors in velocity result in a large volume of error space unlike Sextans, for example, which is the least constrained dwarf in velocity but lies within 100 kpc of the Milky Way and will therefore lie closer to any planar solution due to its proximity to the centre. Previous work has shown that possibly only four dwarfs truly belong to a planar structure: Sculptor, Draco, Sextans, and Carina, three of which have proper motions that are the most poorly constrained. While an alignment of four dwarfs would be unique, it would not threaten to falsify $\Lambda$CDM as an entire VPOS may but maybe instead indicate a large merger event in the past.

The era of large scale surveys and accurate proper motion studies like the Large Synoptic Survey Telescope (LSST) and \textit{Gaia} will inevitably help solve this problem as well as add more VPOS dwarfs to the list of satellites for which we have accurate proper motions. Proper motions of dwarf galaxies are exceedingly difficult to obtain. Without these measurements, it is difficult to distinguish between dwarf galaxies located in a coherent structure and dwarfs that happen to align momentarily. However, as surveys and observations continue to improve we are able to detect more dwarf galaxies around neighboring hosts and may begin to infer the presence of more dwarf planar structures. When observing multiple systems for the same phenomenon, one must consider the ``look elsewhere effect'' which explains that expanding your search area includes increasing your chances of seeing more statistical fluctuations \citep[e.g.][]{Way:2012aa}. In other words, observing multiple significant events is not as unlikely as seeing just one significant event when you continue to broaden your search. This effect can be quantified through the use of the trials factor which defines the probability of falsely defining a distribution as significant by the sum of the probabilities.

\begin{equation}
\text{Prob(false +)} = 1 - (1 - p)^N
\end{equation}

Where $p$ is the probability of falsely judging a distribution as significant and $N$ is the number of trials. For our case, the VPOS is located 2.03$\sigma$ away from the mean of Fig. \ref{delta_plot} (or $p=0.0424$). \citet{Ibata:2013aa} and \citet{Conn:2013aa} showed that 15 of the 27 known dwarfs around M31 are aligned in a plane of thickness, $D_{\text{rms}}=12.34$ kpc or $\delta=0.12$. This means that the plane around M31 is located 3.43$\sigma$ away from the mean (or $p=0.0006$). Here, $p$ represents the number of simulated planes thinner than the observed distribution over the total number of simulated planes. Therefore, if we were to judge these results as significant alignments then the probability that one or both of them are actually false positives is $\text{Prob(false +)}=0.043 = 4.3$ per cent.

\section{Conclusions}
Our main results and conclusions are as follows:

\begin{itemize}

\item We have analysed the Milky Way distribution of classical dwarf galaxies and compared it to distributions of sub-haloes in CDM simulations. Plane fitting was performed using PCA. We compared the Milky Way distribution to the VLII \citep{Diemand:2007aa} and ELVIS \citep{Garrison-Kimmel:2014aa} dark matter-only simulations. Fig. \ref{delta_plot} shows that most solutions from simulations form relatively poor planar structures with a normalized thickness of $\delta\sim0.40$. However, the Milky Way alignment is in rough agreement with this value, lying less than 2$\sigma$ away from the mean, leading to a probability of drawing a more planar distribution from a simulation to about 1 in 24. When only the 50 largest sub-haloes are considered, the probability of drawing a more planar distribution from a simulation increases to about 1 in 10.

\item New proper motion data from \citet{Piatek:2016aa} have significantly improved the accuracy of the proper motion measurements for Leo II. We have presented an orbital analysis of Leo II and find that it has spent much of its time at large distances, making at most one pericentre passage within the last 2.5 Gyr. When this knowledge is combined with information about its star formation history, it appears that Leo II may be on first infall on to the Milky Way halo. Therefore, we argue that Leo II should be excluded from VPOS analyses. 

\item Due to the distant radial location, high velocity, and star formation history of Leo I, it appears that it too is not part of a long-lived structure.

\item When plane-fitting analysis was performed on the remaining classical dwarfs after Leo I and Leo II were removed, a very similar plane was found. This indicates that although Leo I and Leo II are located distantly compared to the other dwarfs, they alone do not drive the polar fit of the VPOS.

\item Through the use of the test particle code developed by \citet{Chang:2011aa}, we integrated the orbits of the dwarfs backward in time to analyse the stability of the VPOS. We found that the structure became less significant over time. This result was seen for all the subsets considered: all 11 classical dwarfs, Leo I and Leo II removed, and dwarfs with aligned orbital angular momentum vectors.

\item We found that adding errors to the components of velocity and radius inevitably led to a loss of coherent structure. For errors up to 100 per cent in each velocity component, coherence was maintained after 0.5 Gyr; for errors greater than 50 per cent, coherence was lost after 0.75 Gyr; and for errors greater than 30 per cent, coherence was lost after 1 Gyr. This shows that eight dwarfs have errors that are constrained enough that orbit integrations can be trusted for up to 0.5 Gyr. Those dwarfs are Sagittarius, LMC, SMC, Draco, Sextans, Fornax, Leo II, and Leo I. When the orbits of these dwarfs are integrated backward in time, they \emph{also} display a loss of significance well within 1 Gyr. Thus, we find that the VPOS is not a stable structure that maintains coherence or significance.\footnote{The authors would like to recognize the work of \citet{Fernando:2017aa} who found that Milky Way/M31 planes are unstable in general; this paper was submitted while our paper was in the review process.}

\item The three remaining dwarfs that have error estimates too large to be trustworthy include Ursa Minor, Sculptor, and Carina. Only three dwarfs have proper motion measurements that are accurate enough to be trusted in orbit integrations beyond 1 Gyr: Sagittarius, LMC, and Draco. The other remaining dwarfs (SMC, Sextans, Fornax, Leo II, and Leo I) suffer from at least one very uncertain velocity component. If the errors on the proper motions of the dwarfs were constrained to a level of $\leq$30 per cent, our conclusion on the stability of the VPOS would be further bolstered.

\end{itemize}

\section*{Acknowledgements}

We would like to thank the referee for helpful suggestions, Chris McKee, Eric Blackman, Philip Chang, Tad Pryor, Tony Sohn, and Daniel Wysocki for helpful discussions, and Research Computing at Rochester Institute of Technology for the use of their cluster. We would also like to thank Shea Garrison-Kimmel and the ELVIS collaboration for making their simulations publicly available. SC and AL were supported by National Science Foundation grant No. 1517488.




\bibliographystyle{mnras}

\bibliography{vposbib}

\begin{thebibliography}{}
\makeatletter
\relax
\def\mn@urlcharsother{\let\do\@makeother \do\$\do\&\do\#\do\^\do\_\do\%\do\~}
\def\mn@doi{\begingroup\mn@urlcharsother \@ifnextchar [ {\mn@doi@}
  {\mn@doi@[]}}
\def\mn@doi@[#1]#2{\def\@tempa{#1}\ifx\@tempa\@empty \href
  {http://dx.doi.org/#2} {doi:#2}\else \href {http://dx.doi.org/#2} {#1}\fi
  \endgroup}
\def\mn@eprint#1#2{\mn@eprint@#1:#2::\@nil}
\def\mn@eprint@arXiv#1{\href {http://arxiv.org/abs/#1} {{\tt arXiv:#1}}}
\def\mn@eprint@dblp#1{\href {http://dblp.uni-trier.de/rec/bibtex/#1.xml}
  {dblp:#1}}
\def\mn@eprint@#1:#2:#3:#4\@nil{\def\@tempa {#1}\def\@tempb {#2}\def\@tempc
  {#3}\ifx \@tempc \@empty \let \@tempc \@tempb \let \@tempb \@tempa \fi \ifx
  \@tempb \@empty \def\@tempb {arXiv}\fi \@ifundefined
  {mn@eprint@\@tempb}{\@tempb:\@tempc}{\expandafter \expandafter \csname
  mn@eprint@\@tempb\endcsname \expandafter{\@tempc}}}

\bibitem[\protect\citeauthoryear{{Ahmed}, {Brooks}  \& {Christensen}}{{Ahmed}
  et~al.}{2016}]{Ahmed:2016aa}
{Ahmed} S.~H.,  {Brooks} A.~M.,   {Christensen} C.~R.,  2016, preprint, \href
  {http://adsabs.harvard.edu/abs/2016arXiv161003077A} {} (\mn@eprint {arXiv}
  {1610.03077})

\bibitem[\protect\citeauthoryear{{Besla}, {Kallivayalil}, {Hernquist},
  {Robertson}, {Cox}, {van der Marel}  \& {Alcock}}{{Besla}
  et~al.}{2007}]{Besla:2007aa}
{Besla} G.,  {Kallivayalil} N.,  {Hernquist} L.,  {Robertson} B.,  {Cox} T.~J.,
   {van der Marel} R.~P.,   {Alcock} C.,  2007, \mn@doi [\apj]
  {10.1086/521385}, \href {http://adsabs.harvard.edu/abs/2007ApJ...668..949B}
  {668, 949}

\bibitem[\protect\citeauthoryear{{Besla}, {Kallivayalil}, {Hernquist}, {van der
  Marel}, {Cox}  \& {Kere{\v s}}}{{Besla} et~al.}{2010}]{Besla:2010aa}
{Besla} G.,  {Kallivayalil} N.,  {Hernquist} L.,  {van der Marel} R.~P.,  {Cox}
  T.~J.,   {Kere{\v s}} D.,  2010, \mn@doi [\apjl]
  {10.1088/2041-8205/721/2/L97}, \href
  {http://adsabs.harvard.edu/abs/2010ApJ...721L..97B} {721, L97}

\bibitem[\protect\citeauthoryear{{Bovy}, {Bahmanyar}, {Fritz}  \&
  {Kallivayalil}}{{Bovy} et~al.}{2016}]{Bovy:2016aa}
{Bovy} J.,  {Bahmanyar} A.,  {Fritz} T.~K.,   {Kallivayalil} N.,  2016, \mn@doi
  [\apj] {10.3847/1538-4357/833/1/31}, \href
  {http://adsabs.harvard.edu/abs/2016ApJ...833...31B} {833, 31}

\bibitem[\protect\citeauthoryear{{Boylan-Kolchin}, {Bullock}  \&
  {Kaplinghat}}{{Boylan-Kolchin} et~al.}{2011}]{Boylan-Kolchin:2011aa}
{Boylan-Kolchin} M.,  {Bullock} J.~S.,   {Kaplinghat} M.,  2011, \mn@doi
  [\mnras] {10.1111/j.1745-3933.2011.01074.x}, \href
  {http://adsabs.harvard.edu/abs/2011MNRAS.415L..40B} {415, L40}

\bibitem[\protect\citeauthoryear{{Boylan-Kolchin}, {Bullock}, {Sohn}, {Besla}
  \& {van der Marel}}{{Boylan-Kolchin} et~al.}{2013}]{Boylan-Kolchin:2013aa}
{Boylan-Kolchin} M.,  {Bullock} J.~S.,  {Sohn} S.~T.,  {Besla} G.,   {van der
  Marel} R.~P.,  2013, \mn@doi [\apj] {10.1088/0004-637X/768/2/140}, \href
  {http://adsabs.harvard.edu/abs/2013ApJ...768..140B} {768, 140}

\bibitem[\protect\citeauthoryear{{Buck}, {Macci{\`o}}  \& {Dutton}}{{Buck}
  et~al.}{2015}]{Buck:2015aa}
{Buck} T.,  {Macci{\`o}} A.~V.,   {Dutton} A.~A.,  2015, \mn@doi [\apj]
  {10.1088/0004-637X/809/1/49}, \href
  {http://adsabs.harvard.edu/abs/2015ApJ...809...49B} {809, 49}

\bibitem[\protect\citeauthoryear{{Buck}, {Dutton}  \& {Macci{\`o}}}{{Buck}
  et~al.}{2016}]{Buck:2016aa}
{Buck} T.,  {Dutton} A.~A.,   {Macci{\`o}} A.~V.,  2016, \mn@doi [\mnras]
  {10.1093/mnras/stw1232}, \href
  {http://adsabs.harvard.edu/abs/2016MNRAS.460.4348B} {460, 4348}

\bibitem[\protect\citeauthoryear{{Casetti-Dinescu} \&
  {Girard}}{{Casetti-Dinescu} \& {Girard}}{2016}]{Casetti-Dinescu:2016aa}
{Casetti-Dinescu} D.~I.,  {Girard} T.~M.,  2016, \mn@doi [\mnras]
  {10.1093/mnras/stw1337}, \href
  {http://adsabs.harvard.edu/abs/2016MNRAS.461..271C} {461, 271}

\bibitem[\protect\citeauthoryear{{Cautun}, {Bose}, {Frenk}, {Guo}, {Han},
  {Hellwing}, {Sawala}  \& {Wang}}{{Cautun} et~al.}{2015}]{Cautun:2015aa}
{Cautun} M.,  {Bose} S.,  {Frenk} C.~S.,  {Guo} Q.,  {Han} J.,  {Hellwing}
  W.~A.,  {Sawala} T.,   {Wang} W.,  2015, \mn@doi [\mnras]
  {10.1093/mnras/stv1557}, \href
  {http://adsabs.harvard.edu/abs/2015MNRAS.452.3838C} {452, 3838}

\bibitem[\protect\citeauthoryear{{Chakrabarti} \& {Blitz}}{{Chakrabarti} \&
  {Blitz}}{2009}]{Chakrabarti:2009aa}
{Chakrabarti} S.,  {Blitz} L.,  2009, \mn@doi [\mnras]
  {10.1111/j.1745-3933.2009.00735.x}, \href
  {http://adsabs.harvard.edu/abs/2009MNRAS.399L.118C} {399, L118}

\bibitem[\protect\citeauthoryear{{Chakrabarti} \& {Blitz}}{{Chakrabarti} \&
  {Blitz}}{2011}]{Chakrabarti:2011aa}
{Chakrabarti} S.,  {Blitz} L.,  2011, \mn@doi [\apj]
  {10.1088/0004-637X/731/1/40}, \href
  {http://adsabs.harvard.edu/abs/2011ApJ...731...40C} {731, 40}

\bibitem[\protect\citeauthoryear{{Chakrabarti}, {Quillen}, {Chang}  \&
  {Merritt}}{{Chakrabarti} et~al.}{2014}]{Chakrabarti:2014aa}
{Chakrabarti} S.,  {Quillen} A.,  {Chang} P.,   {Merritt} D.,  2014, preprint,
  \href {http://adsabs.harvard.edu/abs/2014arXiv1401.4182C} {} (\mn@eprint
  {arXiv} {1401.4182})

\bibitem[\protect\citeauthoryear{{Chakrabarti}, {Saito}, {Quillen}, {Gran},
  {Klein}  \& {Blitz}}{{Chakrabarti} et~al.}{2015}]{Chakrabarti:2015aa}
{Chakrabarti} S.,  {Saito} R.,  {Quillen} A.,  {Gran} F.,  {Klein} C.,
  {Blitz} L.,  2015, \mn@doi [\apjl] {10.1088/2041-8205/802/1/L4}, \href
  {http://adsabs.harvard.edu/abs/2015ApJ...802L...4C} {802, L4}

\bibitem[\protect\citeauthoryear{{Chakrabarti} et~al.,}{{Chakrabarti}
  et~al.}{2016}]{Chakrabarti:2016aa}
{Chakrabarti} S.,  et~al., 2016, preprint, \href
  {http://adsabs.harvard.edu/abs/2016arXiv160103381C} {} (\mn@eprint {arXiv}
  {1601.03381})

\bibitem[\protect\citeauthoryear{{Chang} \& {Chakrabarti}}{{Chang} \&
  {Chakrabarti}}{2011}]{Chang:2011aa}
{Chang} P.,  {Chakrabarti} S.,  2011, \mn@doi [\mnras]
  {10.1111/j.1365-2966.2011.19071.x}, \href
  {http://adsabs.harvard.edu/abs/2011MNRAS.416..618C} {416, 618}

\bibitem[\protect\citeauthoryear{{Collins} et~al.,}{{Collins}
  et~al.}{2015}]{Collins:2015aa}
{Collins} M.~L.~M.,  et~al., 2015, \mn@doi [\apjl]
  {10.1088/2041-8205/799/1/L13}, \href
  {http://adsabs.harvard.edu/abs/2015ApJ...799L..13C} {799, L13}

\bibitem[\protect\citeauthoryear{{Conn} et~al.,}{{Conn}
  et~al.}{2013}]{Conn:2013aa}
{Conn} A.~R.,  et~al., 2013, \mn@doi [\apj] {10.1088/0004-637X/766/2/120},
  \href {http://adsabs.harvard.edu/abs/2013ApJ...766..120C} {766, 120}

\bibitem[\protect\citeauthoryear{{Deason}, {Belokurov}, {Evans}  \&
  {An}}{{Deason} et~al.}{2012}]{Deason:2012aa}
{Deason} A.~J.,  {Belokurov} V.,  {Evans} N.~W.,   {An} J.,  2012, \mn@doi
  [\mnras] {10.1111/j.1745-3933.2012.01283.x}, \href
  {http://adsabs.harvard.edu/abs/2012MNRAS.424L..44D} {424, L44}

\bibitem[\protect\citeauthoryear{{Debattista}, {Moore}, {Quinn}, {Kazantzidis},
  {Maas}, {Mayer}, {Read}  \& {Stadel}}{{Debattista}
  et~al.}{2008}]{Debattista:2008aa}
{Debattista} V.~P.,  {Moore} B.,  {Quinn} T.,  {Kazantzidis} S.,  {Maas} R.,
  {Mayer} L.,  {Read} J.,   {Stadel} J.,  2008, \mn@doi [\apj]
  {10.1086/587977}, \href {http://adsabs.harvard.edu/abs/2008ApJ...681.1076D}
  {681, 1076}

\bibitem[\protect\citeauthoryear{{Debattista}, {Ro{\v s}kar}, {Valluri},
  {Quinn}, {Moore}  \& {Wadsley}}{{Debattista}
  et~al.}{2013}]{Debattista:2013aa}
{Debattista} V.~P.,  {Ro{\v s}kar} R.,  {Valluri} M.,  {Quinn} T.,  {Moore} B.,
    {Wadsley} J.,  2013, \mn@doi [\mnras] {10.1093/mnras/stt1217}, \href
  {http://adsabs.harvard.edu/abs/2013MNRAS.434.2971D} {434, 2971}

\bibitem[\protect\citeauthoryear{{Deg} \& {Widrow}}{{Deg} \&
  {Widrow}}{2014}]{Deg:2014aa}
{Deg} N.,  {Widrow} L.,  2014, \mn@doi [\mnras] {10.1093/mnras/stu132}, \href
  {http://adsabs.harvard.edu/abs/2014MNRAS.439.2678D} {439, 2678}

\bibitem[\protect\citeauthoryear{{Diemand}, {Kuhlen}  \& {Madau}}{{Diemand}
  et~al.}{2007}]{Diemand:2007aa}
{Diemand} J.,  {Kuhlen} M.,   {Madau} P.,  2007, \mn@doi [\apj]
  {10.1086/510736}, \href {http://adsabs.harvard.edu/abs/2007ApJ...657..262D}
  {657, 262}

\bibitem[\protect\citeauthoryear{{Fernando}, {Arias}, {Guglielmo}, {Lewis},
  {Ibata}  \& {Power}}{{Fernando} et~al.}{2017}]{Fernando:2017aa}
{Fernando} N.,  {Arias} V.,  {Guglielmo} M.,  {Lewis} G.~F.,  {Ibata} R.~A.,
  {Power} C.,  2017, \mn@doi [\mnras] {10.1093/mnras/stw2694}, \href
  {http://adsabs.harvard.edu/abs/2017MNRAS.465..641F} {465, 641}

\bibitem[\protect\citeauthoryear{{Foot} \& {Vagnozzi}}{{Foot} \&
  {Vagnozzi}}{2016}]{Foot:2016aa}
{Foot} R.,  {Vagnozzi} S.,  2016, \mn@doi [\jcap]
  {10.1088/1475-7516/2016/07/013}, \href
  {http://adsabs.harvard.edu/abs/2016JCAP...07..013F} {7, 013}

\bibitem[\protect\citeauthoryear{{Garrison-Kimmel}, {Boylan-Kolchin}, {Bullock}
   \& {Lee}}{{Garrison-Kimmel} et~al.}{2014}]{Garrison-Kimmel:2014aa}
{Garrison-Kimmel} S.,  {Boylan-Kolchin} M.,  {Bullock} J.~S.,   {Lee} K.,
  2014, \mn@doi [\mnras] {10.1093/mnras/stt2377}, \href
  {http://adsabs.harvard.edu/abs/2014MNRAS.438.2578G} {438, 2578}

\bibitem[\protect\citeauthoryear{{Genel} et~al.,}{{Genel}
  et~al.}{2014}]{Genel:2014aa}
{Genel} S.,  et~al., 2014, \mn@doi [\mnras] {10.1093/mnras/stu1654}, \href
  {http://adsabs.harvard.edu/abs/2014MNRAS.445..175G} {445, 175}

\bibitem[\protect\citeauthoryear{{Guedes}, {Callegari}, {Madau}  \&
  {Mayer}}{{Guedes} et~al.}{2011}]{Guedes:2011aa}
{Guedes} J.,  {Callegari} S.,  {Madau} P.,   {Mayer} L.,  2011, \mn@doi [\apj]
  {10.1088/0004-637X/742/2/76}, \href
  {http://adsabs.harvard.edu/abs/2011ApJ...742...76G} {742, 76}

\bibitem[\protect\citeauthoryear{{Haider}, {Steinhauser}, {Vogelsberger},
  {Genel}, {Springel}, {Torrey}  \& {Hernquist}}{{Haider}
  et~al.}{2016}]{Haider:2016aa}
{Haider} M.,  {Steinhauser} D.,  {Vogelsberger} M.,  {Genel} S.,  {Springel}
  V.,  {Torrey} P.,   {Hernquist} L.,  2016, \mn@doi [\mnras]
  {10.1093/mnras/stw077}, \href
  {http://adsabs.harvard.edu/abs/2016MNRAS.457.3024H} {457, 3024}

\bibitem[\protect\citeauthoryear{{Hammer}, {Yang}, {Fouquet}, {Pawlowski},
  {Kroupa}, {Puech}, {Flores}  \& {Wang}}{{Hammer}
  et~al.}{2013}]{Hammer:2013aa}
{Hammer} F.,  {Yang} Y.,  {Fouquet} S.,  {Pawlowski} M.~S.,  {Kroupa} P.,
  {Puech} M.,  {Flores} H.,   {Wang} J.,  2013, \mn@doi [\mnras]
  {10.1093/mnras/stt435}, \href
  {http://adsabs.harvard.edu/abs/2013MNRAS.431.3543H} {431, 3543}

\bibitem[\protect\citeauthoryear{{Hernquist}}{{Hernquist}}{1990}]{Hernquist:1990aa}
{Hernquist} L.,  1990, \mn@doi [\apj] {10.1086/168845}, \href
  {http://adsabs.harvard.edu/abs/1990ApJ...356..359H} {356, 359}

\bibitem[\protect\citeauthoryear{{Homma} et~al.,}{{Homma}
  et~al.}{2016}]{Homma:2016aa}
{Homma} D.,  et~al., 2016, \mn@doi [\apj] {10.3847/0004-637X/832/1/21}, \href
  {http://adsabs.harvard.edu/abs/2016ApJ...832...21H} {832, 21}

\bibitem[\protect\citeauthoryear{{Ibata} et~al.,}{{Ibata}
  et~al.}{2013}]{Ibata:2013aa}
{Ibata} R.~A.,  et~al., 2013, \mn@doi [\nat] {10.1038/nature11717}, \href
  {http://adsabs.harvard.edu/abs/2013Natur.493...62I} {493, 62}

\bibitem[\protect\citeauthoryear{{Johnston}, {Majewski}, {Siegel}, {Reid}  \&
  {Kunkel}}{{Johnston} et~al.}{1999}]{Johnston:1999aa}
{Johnston} K.~V.,  {Majewski} S.~R.,  {Siegel} M.~H.,  {Reid} I.~N.,   {Kunkel}
  W.~E.,  1999, \mn@doi [\aj] {10.1086/301037}, \href
  {http://adsabs.harvard.edu/abs/1999AJ....118.1719J} {118, 1719}

\bibitem[\protect\citeauthoryear{{Kallivayalil}, {van der Marel}, {Besla},
  {Anderson}  \& {Alcock}}{{Kallivayalil} et~al.}{2013}]{Kallivayalil:2013aa}
{Kallivayalil} N.,  {van der Marel} R.~P.,  {Besla} G.,  {Anderson} J.,
  {Alcock} C.,  2013, \mn@doi [\apj] {10.1088/0004-637X/764/2/161}, \href
  {http://adsabs.harvard.edu/abs/2013ApJ...764..161K} {764, 161}

\bibitem[\protect\citeauthoryear{{Kang}, {Mao}, {Gao}  \& {Jing}}{{Kang}
  et~al.}{2005}]{Kang:2005aa}
{Kang} X.,  {Mao} S.,  {Gao} L.,   {Jing} Y.~P.,  2005, \mn@doi [\aap]
  {10.1051/0004-6361:20052675}, \href
  {http://adsabs.harvard.edu/abs/2005A%26A...437..383K} {437, 383}

\bibitem[\protect\citeauthoryear{{Klypin}, {Kravtsov}, {Valenzuela}  \&
  {Prada}}{{Klypin} et~al.}{1999}]{Klypin:1999aa}
{Klypin} A.,  {Kravtsov} A.~V.,  {Valenzuela} O.,   {Prada} F.,  1999, \mn@doi
  [\apj] {10.1086/307643}, \href
  {http://adsabs.harvard.edu/abs/1999ApJ...522...82K} {522, 82}

\bibitem[\protect\citeauthoryear{{Kravtsov}, {Gnedin}  \& {Klypin}}{{Kravtsov}
  et~al.}{2004}]{Kravtsov:2004aa}
{Kravtsov} A.~V.,  {Gnedin} O.~Y.,   {Klypin} A.~A.,  2004, \mn@doi [\apj]
  {10.1086/421322}, \href {http://adsabs.harvard.edu/abs/2004ApJ...609..482K}
  {609, 482}

\bibitem[\protect\citeauthoryear{{Kroupa}, {Theis}  \& {Boily}}{{Kroupa}
  et~al.}{2005}]{Kroupa:2005aa}
{Kroupa} P.,  {Theis} C.,   {Boily} C.~M.,  2005, \mn@doi [\aap]
  {10.1051/0004-6361:20041122}, \href
  {http://adsabs.harvard.edu/abs/2005A%26A...431..517K} {431, 517}

\bibitem[\protect\citeauthoryear{{Kroupa} et~al.,}{{Kroupa}
  et~al.}{2010}]{Kroupa:2010aa}
{Kroupa} P.,  et~al., 2010, \mn@doi [\aap] {10.1051/0004-6361/201014892}, \href
  {http://adsabs.harvard.edu/abs/2010A%26A...523A..32K} {523, A32}

\bibitem[\protect\citeauthoryear{{Kunkel} \& {Demers}}{{Kunkel} \&
  {Demers}}{1976}]{Kunkel:1976aa}
{Kunkel} W.~E.,  {Demers} S.,  1976, in {Dickens} R.~J.,  {Perry} J.~E.,
  {Smith} F.~G.,   {King} I.~R.,  eds,  Royal Greenwich Observatory Bulletins
  Vol. 182, The Galaxy and the Local Group. p.~241

\bibitem[\protect\citeauthoryear{{Larson} et~al.,}{{Larson}
  et~al.}{2011}]{Larson:2011aa}
{Larson} D.,  et~al., 2011, \mn@doi [\apjs] {10.1088/0067-0049/192/2/16}, \href
  {http://adsabs.harvard.edu/abs/2011ApJS..192...16L} {192, 16}

\bibitem[\protect\citeauthoryear{{Law} \& {Majewski}}{{Law} \&
  {Majewski}}{2010}]{Law:2010aa}
{Law} D.~R.,  {Majewski} S.~R.,  2010, \mn@doi [\apj]
  {10.1088/0004-637X/714/1/229}, \href
  {http://adsabs.harvard.edu/abs/2010ApJ...714..229L} {714, 229}

\bibitem[\protect\citeauthoryear{{L{\'e}pine}, {Koch}, {Rich}  \&
  {Kuijken}}{{L{\'e}pine} et~al.}{2011}]{Lepine:2011aa}
{L{\'e}pine} S.,  {Koch} A.,  {Rich} R.~M.,   {Kuijken} K.,  2011, \mn@doi
  [\apj] {10.1088/0004-637X/741/2/100}, \href
  {http://adsabs.harvard.edu/abs/2011ApJ...741..100L} {741, 100}

\bibitem[\protect\citeauthoryear{{Libeskind}, {Cole}, {Frenk}, {Okamoto}  \&
  {Jenkins}}{{Libeskind} et~al.}{2007}]{Libeskind:2007aa}
{Libeskind} N.~I.,  {Cole} S.,  {Frenk} C.~S.,  {Okamoto} T.,   {Jenkins} A.,
  2007, \mn@doi [\mnras] {10.1111/j.1365-2966.2006.11205.x}, \href
  {http://adsabs.harvard.edu/abs/2007MNRAS.374...16L} {374, 16}

\bibitem[\protect\citeauthoryear{{Libeskind}, {Knebe}, {Hoffman}  \&
  {Gottl{\"o}ber}}{{Libeskind} et~al.}{2014}]{Libeskind:2014aa}
{Libeskind} N.~I.,  {Knebe} A.,  {Hoffman} Y.,   {Gottl{\"o}ber} S.,  2014,
  \mn@doi [\mnras] {10.1093/mnras/stu1216}, \href
  {http://adsabs.harvard.edu/abs/2014MNRAS.443.1274L} {443, 1274}

\bibitem[\protect\citeauthoryear{{Libeskind}, {Hoffman}, {Tully}, {Courtois},
  {Pomar{\`e}de}, {Gottl{\"o}ber}  \& {Steinmetz}}{{Libeskind}
  et~al.}{2015}]{Libeskind:2015aa}
{Libeskind} N.~I.,  {Hoffman} Y.,  {Tully} R.~B.,  {Courtois} H.~M.,
  {Pomar{\`e}de} D.,  {Gottl{\"o}ber} S.,   {Steinmetz} M.,  2015, \mn@doi
  [\mnras] {10.1093/mnras/stv1302}, \href
  {http://adsabs.harvard.edu/abs/2015MNRAS.452.1052L} {452, 1052}

\bibitem[\protect\citeauthoryear{{Loebman}, {Ivezi{\'c}}, {Quinn}, {Governato},
  {Brooks}, {Christensen}  \& {Juri{\'c}}}{{Loebman}
  et~al.}{2012}]{Loebman:2012aa}
{Loebman} S.~R.,  {Ivezi{\'c}} {\v Z}.,  {Quinn} T.~R.,  {Governato} F.,
  {Brooks} A.~M.,  {Christensen} C.~R.,   {Juri{\'c}} M.,  2012, \mn@doi
  [\apjl] {10.1088/2041-8205/758/1/L23}, \href
  {http://adsabs.harvard.edu/abs/2012ApJ...758L..23L} {758, L23}

\bibitem[\protect\citeauthoryear{{Lynden-Bell}}{{Lynden-Bell}}{1976}]{Lynden-Bell:1976aa}
{Lynden-Bell} D.,  1976, \mnras, \href
  {http://adsabs.harvard.edu/abs/1976MNRAS.174..695L} {174, 695}

\bibitem[\protect\citeauthoryear{{Majewski}, {Skrutskie}, {Weinberg}  \&
  {Ostheimer}}{{Majewski} et~al.}{2003}]{Majewski:2003aa}
{Majewski} S.~R.,  {Skrutskie} M.~F.,  {Weinberg} M.~D.,   {Ostheimer} J.~C.,
  2003, \mn@doi [\apj] {10.1086/379504}, \href
  {http://adsabs.harvard.edu/abs/2003ApJ...599.1082M} {599, 1082}

\bibitem[\protect\citeauthoryear{{Massari}, {Bellini}, {Ferraro}, {van der
  Marel}, {Anderson}, {Dalessandro}  \& {Lanzoni}}{{Massari}
  et~al.}{2013}]{Massari:2013aa}
{Massari} D.,  {Bellini} A.,  {Ferraro} F.~R.,  {van der Marel} R.~P.,
  {Anderson} J.,  {Dalessandro} E.,   {Lanzoni} B.,  2013, \mn@doi [\apj]
  {10.1088/0004-637X/779/1/81}, \href
  {http://adsabs.harvard.edu/abs/2013ApJ...779...81M} {779, 81}

\bibitem[\protect\citeauthoryear{{Mateo}}{{Mateo}}{1998}]{Mateo:1998aa}
{Mateo} M.~L.,  1998, \mn@doi [\araa] {10.1146/annurev.astro.36.1.435}, \href
  {http://adsabs.harvard.edu/abs/1998ARA%26A..36..435M} {36, 435}

\bibitem[\protect\citeauthoryear{{Mateo}, {Olszewski}  \& {Walker}}{{Mateo}
  et~al.}{2008}]{Mateo:2008aa}
{Mateo} M.,  {Olszewski} E.~W.,   {Walker} M.~G.,  2008, \mn@doi [\apj]
  {10.1086/522326}, \href {http://adsabs.harvard.edu/abs/2008ApJ...675..201M}
  {675, 201}

\bibitem[\protect\citeauthoryear{{McCall}}{{McCall}}{2014}]{McCall:2014aa}
{McCall} M.~L.,  2014, \mn@doi [\mnras] {10.1093/mnras/stu199}, \href
  {http://adsabs.harvard.edu/abs/2014MNRAS.440..405M} {440, 405}

\bibitem[\protect\citeauthoryear{{McConnachie}}{{McConnachie}}{2012}]{McConnachie:2012aa}
{McConnachie} A.~W.,  2012, \mn@doi [\aj] {10.1088/0004-6256/144/1/4}, \href
  {http://adsabs.harvard.edu/abs/2012AJ....144....4M} {144, 4}

\bibitem[\protect\citeauthoryear{{Metz}, {Kroupa}  \& {Jerjen}}{{Metz}
  et~al.}{2007}]{Metz:2007aa}
{Metz} M.,  {Kroupa} P.,   {Jerjen} H.,  2007, \mn@doi [\mnras]
  {10.1111/j.1365-2966.2006.11228.x}, \href
  {http://adsabs.harvard.edu/abs/2007MNRAS.374.1125M} {374, 1125}

\bibitem[\protect\citeauthoryear{{Moore}, {Ghigna}, {Governato}, {Lake},
  {Quinn}, {Stadel}  \& {Tozzi}}{{Moore} et~al.}{1999}]{Moore:1999aa}
{Moore} B.,  {Ghigna} S.,  {Governato} F.,  {Lake} G.,  {Quinn} T.,  {Stadel}
  J.,   {Tozzi} P.,  1999, \mn@doi [\apjl] {10.1086/312287}, \href
  {http://adsabs.harvard.edu/abs/1999ApJ...524L..19M} {524, L19}

\bibitem[\protect\citeauthoryear{{Navarro}, {Frenk}  \& {White}}{{Navarro}
  et~al.}{1997}]{Navarro:1997aa}
{Navarro} J.~F.,  {Frenk} C.~S.,   {White} S.~D.~M.,  1997, \apj, \href
  {http://adsabs.harvard.edu/abs/1997ApJ...490..493N} {490, 493}

\bibitem[\protect\citeauthoryear{{Palma}, {Majewski}  \& {Johnston}}{{Palma}
  et~al.}{2002}]{Palma:2002aa}
{Palma} C.,  {Majewski} S.~R.,   {Johnston} K.~V.,  2002, \mn@doi [\apj]
  {10.1086/324137}, \href {http://adsabs.harvard.edu/abs/2002ApJ...564..736P}
  {564, 736}

\bibitem[\protect\citeauthoryear{{Pawlowski} \& {Kroupa}}{{Pawlowski} \&
  {Kroupa}}{2013}]{Pawlowski:2013aa}
{Pawlowski} M.~S.,  {Kroupa} P.,  2013, \mn@doi [\mnras]
  {10.1093/mnras/stt1429}, \href
  {http://adsabs.harvard.edu/abs/2013MNRAS.435.2116P} {435, 2116}

\bibitem[\protect\citeauthoryear{{Pawlowski} \& {McGaugh}}{{Pawlowski} \&
  {McGaugh}}{2014}]{Pawlowski:2014aa}
{Pawlowski} M.~S.,  {McGaugh} S.~S.,  2014, \mn@doi [\apjl]
  {10.1088/2041-8205/789/1/L24}, \href
  {http://adsabs.harvard.edu/abs/2014ApJ...789L..24P} {789, L24}

\bibitem[\protect\citeauthoryear{{Pawlowski} et~al.,}{{Pawlowski}
  et~al.}{2014}]{Pawlowski:2014ab}
{Pawlowski} M.~S.,  et~al., 2014, \mn@doi [\mnras] {10.1093/mnras/stu1005},
  \href {http://adsabs.harvard.edu/abs/2014MNRAS.442.2362P} {442, 2362}

\bibitem[\protect\citeauthoryear{{Pawlowski}, {Famaey}, {Merritt}  \&
  {Kroupa}}{{Pawlowski} et~al.}{2015}]{Pawlowski:2015aa}
{Pawlowski} M.~S.,  {Famaey} B.,  {Merritt} D.,   {Kroupa} P.,  2015, \mn@doi
  [\apj] {10.1088/0004-637X/815/1/19}, \href
  {http://adsabs.harvard.edu/abs/2015ApJ...815...19P} {815, 19}

\bibitem[\protect\citeauthoryear{{Piatek}, {Pryor}, {Bristow}, {Olszewski},
  {Harris}, {Mateo}, {Minniti}  \& {Tinney}}{{Piatek}
  et~al.}{2005}]{Piatek:2005aa}
{Piatek} S.,  {Pryor} C.,  {Bristow} P.,  {Olszewski} E.~W.,  {Harris} H.~C.,
  {Mateo} M.,  {Minniti} D.,   {Tinney} C.~G.,  2005, \mn@doi [\aj]
  {10.1086/430532}, \href {http://adsabs.harvard.edu/abs/2005AJ....130...95P}
  {130, 95}

\bibitem[\protect\citeauthoryear{{Piatek}, {Pryor}, {Bristow}, {Olszewski},
  {Harris}, {Mateo}, {Minniti}  \& {Tinney}}{{Piatek}
  et~al.}{2007}]{Piatek:2007aa}
{Piatek} S.,  {Pryor} C.,  {Bristow} P.,  {Olszewski} E.~W.,  {Harris} H.~C.,
  {Mateo} M.,  {Minniti} D.,   {Tinney} C.~G.,  2007, \mn@doi [\aj]
  {10.1086/510456}, \href {http://adsabs.harvard.edu/abs/2007AJ....133..818P}
  {133, 818}

\bibitem[\protect\citeauthoryear{{Piatek}, {Pryor}  \& {Olszewski}}{{Piatek}
  et~al.}{2016}]{Piatek:2016aa}
{Piatek} S.,  {Pryor} C.,   {Olszewski} E.~W.,  2016, \mn@doi [\aj]
  {10.3847/0004-6256/152/6/166}, \href
  {http://adsabs.harvard.edu/abs/2016AJ....152..166P} {152, 166}

\bibitem[\protect\citeauthoryear{{Price-Whelan} \& {Johnston}}{{Price-Whelan}
  \& {Johnston}}{2013}]{Price-Whelan:2013aa}
{Price-Whelan} A.~M.,  {Johnston} K.~V.,  2013, \mn@doi [\apjl]
  {10.1088/2041-8205/778/1/L12}, \href
  {http://adsabs.harvard.edu/abs/2013ApJ...778L..12P} {778, L12}

\bibitem[\protect\citeauthoryear{{Price-Whelan}, {Hogg}, {Johnston}  \&
  {Hendel}}{{Price-Whelan} et~al.}{2014}]{Price-Whelan:2014aa}
{Price-Whelan} A.~M.,  {Hogg} D.~W.,  {Johnston} K.~V.,   {Hendel} D.,  2014,
  \mn@doi [\apj] {10.1088/0004-637X/794/1/4}, \href
  {http://adsabs.harvard.edu/abs/2014ApJ...794....4P} {794, 4}

\bibitem[\protect\citeauthoryear{{Purcell}, {Bullock}, {Tollerud}, {Rocha}  \&
  {Chakrabarti}}{{Purcell} et~al.}{2011}]{Purcell:2011aa}
{Purcell} C.~W.,  {Bullock} J.~S.,  {Tollerud} E.~J.,  {Rocha} M.,
  {Chakrabarti} S.,  2011, \mn@doi [\nat] {10.1038/nature10417}, \href
  {http://adsabs.harvard.edu/abs/2011Natur.477..301P} {477, 301}

\bibitem[\protect\citeauthoryear{{Randall} \& {Scholtz}}{{Randall} \&
  {Scholtz}}{2015}]{Randall:2015aa}
{Randall} L.,  {Scholtz} J.,  2015, \mn@doi [\jcap]
  {10.1088/1475-7516/2015/09/057}, \href
  {http://adsabs.harvard.edu/abs/2015JCAP...09..057R} {9, 057}

\bibitem[\protect\citeauthoryear{{Rocha}, {Peter}  \& {Bullock}}{{Rocha}
  et~al.}{2012}]{Rocha:2012aa}
{Rocha} M.,  {Peter} A.~H.~G.,   {Bullock} J.,  2012, \mn@doi [\mnras]
  {10.1111/j.1365-2966.2012.21432.x}, \href
  {http://adsabs.harvard.edu/abs/2012MNRAS.425..231R} {425, 231}

\bibitem[\protect\citeauthoryear{{Sales}, {Navarro}, {Abadi}  \&
  {Steinmetz}}{{Sales} et~al.}{2007}]{Sales:2007aa}
{Sales} L.~V.,  {Navarro} J.~F.,  {Abadi} M.~G.,   {Steinmetz} M.,  2007,
  \mn@doi [\mnras] {10.1111/j.1365-2966.2007.12024.x}, \href
  {http://adsabs.harvard.edu/abs/2007MNRAS.379.1464S} {379, 1464}

\bibitem[\protect\citeauthoryear{{Sawala} et~al.,}{{Sawala}
  et~al.}{2014}]{Sawala:2014aa}
{Sawala} T.,  et~al., 2014, preprint, \href
  {http://adsabs.harvard.edu/abs/2014arXiv1412.2748S} {} (\mn@eprint {arXiv}
  {1412.2748})

\bibitem[\protect\citeauthoryear{{Sawala} et~al.,}{{Sawala}
  et~al.}{2016}]{Sawala:2016aa}
{Sawala} T.,  et~al., 2016, \mn@doi [\mnras] {10.1093/mnras/stw145}, \href
  {http://adsabs.harvard.edu/abs/2016MNRAS.457.1931S} {457, 1931}

\bibitem[\protect\citeauthoryear{{Shao}, {Cautun}, {Frenk}, {Gao}, {Crain},
  {Schaller}, {Schaye}  \& {Theuns}}{{Shao} et~al.}{2016}]{Shao:2016aa}
{Shao} S.,  {Cautun} M.,  {Frenk} C.~S.,  {Gao} L.,  {Crain} R.~A.,  {Schaller}
  M.,  {Schaye} J.,   {Theuns} T.,  2016, \mn@doi [\mnras]
  {10.1093/mnras/stw1247}, \href
  {http://adsabs.harvard.edu/abs/2016MNRAS.460.3772S} {460, 3772}

\bibitem[\protect\citeauthoryear{{Shaya} \& {Tully}}{{Shaya} \&
  {Tully}}{2013}]{Shaya:2013aa}
{Shaya} E.~J.,  {Tully} R.~B.,  2013, \mn@doi [\mnras] {10.1093/mnras/stt1714},
  \href {http://adsabs.harvard.edu/abs/2013MNRAS.436.2096S} {436, 2096}

\bibitem[\protect\citeauthoryear{{Sohn}, {Besla}, {van der Marel},
  {Boylan-Kolchin}, {Majewski}  \& {Bullock}}{{Sohn}
  et~al.}{2013}]{Sohn:2013aa}
{Sohn} S.~T.,  {Besla} G.,  {van der Marel} R.~P.,  {Boylan-Kolchin} M.,
  {Majewski} S.~R.,   {Bullock} J.~S.,  2013, \mn@doi [\apj]
  {10.1088/0004-637X/768/2/139}, \href
  {http://adsabs.harvard.edu/abs/2013ApJ...768..139S} {768, 139}

\bibitem[\protect\citeauthoryear{{Spergel} et~al.,}{{Spergel}
  et~al.}{2007}]{Spergel:2007aa}
{Spergel} D.~N.,  et~al., 2007, \mn@doi [\apjs] {10.1086/513700}, \href
  {http://adsabs.harvard.edu/abs/2007ApJS..170..377S} {170, 377}

\bibitem[\protect\citeauthoryear{{Springel}}{{Springel}}{2005}]{Springel:2005aa}
{Springel} V.,  2005, \mn@doi [\mnras] {10.1111/j.1365-2966.2005.09655.x},
  \href {http://adsabs.harvard.edu/abs/2005MNRAS.364.1105S} {364, 1105}

\bibitem[\protect\citeauthoryear{{Stadel}}{{Stadel}}{2001}]{Stadel:2001aa}
{Stadel} J.~G.,  2001, PhD thesis, University of Washington

\bibitem[\protect\citeauthoryear{{Wadsley}, {Stadel}  \& {Quinn}}{{Wadsley}
  et~al.}{2004}]{Wadsley:2004aa}
{Wadsley} J.~W.,  {Stadel} J.,   {Quinn} T.,  2004, \mn@doi [\na]
  {10.1016/j.newast.2003.08.004}, \href
  {http://adsabs.harvard.edu/abs/2004NewA....9..137W} {9, 137}

\bibitem[\protect\citeauthoryear{{Walker}, {Mateo}, {Olszewski},
  {Pe{\~n}arrubia}, {Wyn Evans}  \& {Gilmore}}{{Walker}
  et~al.}{2009}]{Walker:2009aa}
{Walker} M.~G.,  {Mateo} M.,  {Olszewski} E.~W.,  {Pe{\~n}arrubia} J.,  {Wyn
  Evans} N.,   {Gilmore} G.,  2009, \mn@doi [\apj]
  {10.1088/0004-637X/704/2/1274}, \href
  {http://adsabs.harvard.edu/abs/2009ApJ...704.1274W} {704, 1274}

\bibitem[\protect\citeauthoryear{{Wang}, {Frenk}, {Navarro}, {Gao}  \&
  {Sawala}}{{Wang} et~al.}{2012}]{Wang:2012aa}
{Wang} J.,  {Frenk} C.~S.,  {Navarro} J.~F.,  {Gao} L.,   {Sawala} T.,  2012,
  \mn@doi [\mnras] {10.1111/j.1365-2966.2012.21357.x}, \href
  {http://adsabs.harvard.edu/abs/2012MNRAS.424.2715W} {424, 2715}

\bibitem[\protect\citeauthoryear{{Wang}, {Frenk}  \& {Cooper}}{{Wang}
  et~al.}{2013}]{Wang:2013aa}
{Wang} J.,  {Frenk} C.~S.,   {Cooper} A.~P.,  2013, \mn@doi [\mnras]
  {10.1093/mnras/sts442}, \href
  {http://adsabs.harvard.edu/abs/2013MNRAS.429.1502W} {429, 1502}

\bibitem[\protect\citeauthoryear{{Wang}, {Han}, {Cooper}, {Cole}, {Frenk}  \&
  {Lowing}}{{Wang} et~al.}{2015}]{Wang:2015aa}
{Wang} W.,  {Han} J.,  {Cooper} A.~P.,  {Cole} S.,  {Frenk} C.,   {Lowing} B.,
  2015, \mn@doi [\mnras] {10.1093/mnras/stv1647}, \href
  {http://adsabs.harvard.edu/abs/2015MNRAS.453..377W} {453, 377}

\bibitem[\protect\citeauthoryear{{Watkins}, {Evans}  \& {An}}{{Watkins}
  et~al.}{2010}]{Watkins:2010aa}
{Watkins} L.~L.,  {Evans} N.~W.,   {An} J.~H.,  2010, \mn@doi [\mnras]
  {10.1111/j.1365-2966.2010.16708.x}, \href
  {http://adsabs.harvard.edu/abs/2010MNRAS.406..264W} {406, 264}

\bibitem[\protect\citeauthoryear{Way, Scargle, Ali  \& Srivastava}{Way
  et~al.}{2012}]{Way:2012aa}
Way M.~J.,  Scargle J.~D.,  Ali K.~M.,   Srivastava A.~N.,  2012, Advances in
  machine learning and data mining for astronomy.
Chapman \& Hall/CRC data mining and knowledge discovery series, CRC Press, Boca
  Raton, FL, \url {http://opac.inria.fr/record=b1133742}

\bibitem[\protect\citeauthoryear{{Willman}, {Governato}, {Dalcanton}, {Reed}
  \& {Quinn}}{{Willman} et~al.}{2004}]{Willman:2004aa}
{Willman} B.,  {Governato} F.,  {Dalcanton} J.~J.,  {Reed} D.,   {Quinn} T.,
  2004, \mn@doi [\mnras] {10.1111/j.1365-2966.2004.08095.x}, \href
  {http://adsabs.harvard.edu/abs/2004MNRAS.353..639W} {353, 639}

\bibitem[\protect\citeauthoryear{{Yniguez}, {Garrison-Kimmel}, {Boylan-Kolchin}
   \& {Bullock}}{{Yniguez} et~al.}{2014}]{Yniguez:2014aa}
{Yniguez} B.,  {Garrison-Kimmel} S.,  {Boylan-Kolchin} M.,   {Bullock} J.~S.,
  2014, \mn@doi [\mnras] {10.1093/mnras/stt2058}, \href
  {http://adsabs.harvard.edu/abs/2014MNRAS.439...73Y} {439, 73}

\bibitem[\protect\citeauthoryear{{Zentner} \& {Bullock}}{{Zentner} \&
  {Bullock}}{2003}]{Zentner:2003aa}
{Zentner} A.~R.,  {Bullock} J.~S.,  2003, \mn@doi [\apj] {10.1086/378797},
  \href {http://adsabs.harvard.edu/abs/2003ApJ...598...49Z} {598, 49}

\bibitem[\protect\citeauthoryear{{Zentner}, {Kravtsov}, {Gnedin}  \&
  {Klypin}}{{Zentner} et~al.}{2005}]{Zentner:2005aa}
{Zentner} A.~R.,  {Kravtsov} A.~V.,  {Gnedin} O.~Y.,   {Klypin} A.~A.,  2005,
  \mn@doi [\apj] {10.1086/431355}, \href
  {http://adsabs.harvard.edu/abs/2005ApJ...629..219Z} {629, 219}

\bibitem[\protect\citeauthoryear{{Zhao}}{{Zhao}}{1998}]{Zhao:1998aa}
{Zhao} H.,  1998, \mn@doi [\apjl] {10.1086/311413}, \href
  {http://adsabs.harvard.edu/abs/1998ApJ...500L.149Z} {500, L149}

\bibitem[\protect\citeauthoryear{{van der Marel} \& {Kallivayalil}}{{van der
  Marel} \& {Kallivayalil}}{2014}]{van-der-Marel:2014aa}
{van der Marel} R.~P.,  {Kallivayalil} N.,  2014, \mn@doi [\apj]
  {10.1088/0004-637X/781/2/121}, \href
  {http://adsabs.harvard.edu/abs/2014ApJ...781..121V} {781, 121}

\makeatother
\end{thebibliography}


\appendix
\section{Milky Way Dwarf Galaxy Data}

\begin{table*}
\caption{Information of each of the 11 classical dwarf galaxies ranked by distance. Those dwarfs listed in bold have errors that are accurate enough to be trusted in orbit integrations past 0.5 Gyr (see Section 4.1). In this coordinate system, X points from the Sun to the Galactic Centre, Y points in the direction of galactic rotation, and Z points towards the northern galactic pole. References: (1) \citet{Casetti-Dinescu:2016aa} (2) \citet{Kallivayalil:2013aa} (3) \citet{Majewski:2003aa} (4) \citet{Massari:2013aa} (5) \citet{Piatek:2016aa} (6) Pryor priv. comm. (7) \citet{Sohn:2013aa}. All errors from Sohn private communication, except the velocity errors for Sagittarius (Pryor, private communication), Leo II \citep{Piatek:2016aa}, Draco \citep{Casetti-Dinescu:2016aa}, and LMC and SMC \citep{Kallivayalil:2013aa}.}
\begin{center}
{\renewcommand{\arraystretch}{1.5}
\begin{tabular}{ccccccccccc}
\hline\hline
Name			& D (kpc)		& X (kpc)			& Y (kpc)			& Z (kpc)			& V (km s$^{-1}$)		& $V_x$ (km s$^{-1}$)		& $V_y$ (km s$^{-1}$) 	& $V_z$ (km s$^{-1}$)	& Ref.			 \\ \hline
\textbf{Sagittarius} 	& 20.6$\pm$0.5& 19.2$\pm$0.5	&2.70$\pm$0.05	&-6.9$\pm$0.1		& 318$\pm$10	& 235$\pm$4		& -49$\pm$15		& 208$\pm$14	& 3, 4, 6		\\

\textbf{LMC}		& 50$\pm$2	& -1.1$\pm$0.4		& -41$\pm$2		& -28$\pm$1		& 321$\pm$24	& -57$\pm$13		& -226$\pm$15		& 221$\pm$19	& 2, 7			\\

\textbf{SMC}		& 59$\pm$2	& 15.3$\pm$ 0.9	& -37$\pm$2		& -43$\pm$2		& 217$\pm$26	& 19	$\pm$ 18		& -153$\pm$21		& 153$\pm$17	& 2, 7		\\ 

Ursa Minor	& 78$\pm$3	& -22.2$\pm$ 0.8	&52$\pm$3		&54$\pm$3		& 159$\pm$43	& -108$\pm$51		& -15$\pm$34		& -116$\pm$34	& 7		\\

Sculptor	& 85$\pm$1	& -5.3$\pm$0.2		& -9.6$\pm$0.2		& -84$\pm$1		& 248$\pm$39	& -19$\pm$42		& 225$\pm$43		&-102$\pm$5	& 7		\\

\textbf{Draco}		& 93$\pm$4	& -3.5$\pm$0.4		& 76$\pm$5		& 53$\pm$3		& 136$\pm$16	& 95$\pm$18		& -73$\pm$11		& -63$\pm$17	& 1, 7		\\ 

\textbf{Sextans}			& 100$\pm$2	& -40.0$\pm$ 0.9	& -64$\pm$2		&65$\pm$2		& 242$\pm$106& -181$\pm$116	& 114$\pm$98		&114	$\pm$85	& 7		\\

Carina			& 106$\pm$1	& -24.8$\pm$0.3 	& -95$\pm$1		& -39.3$\pm$0.5	& 83$\pm$36	& -73$\pm$38		& 7	$\pm$14		& 38	$\pm$30 	& 7		\\ 

\textbf{Fornax}		& 144$\pm$1	& -40.0$\pm$0.4	& -49.2$\pm$0.5 	& -129$\pm$1		& 178$\pm$20	& -25$\pm$23		& -141$\pm$23		& 106$\pm$11	& 7		\\ 

\textbf{Leo II}		& 236$\pm$7	& -77$\pm$3 		& -58$\pm$2		& 215$\pm$8		& 129$\pm$39& -41$\pm$40		& 116$\pm$41	& 41$\pm$16	& 5, 7		\\ 

\textbf{Leo I}		& 261$\pm$8	& -125$\pm$6		& -121$\pm$6		& 194$\pm$10		& 196$\pm$30	& -168$\pm$32		& -37$\pm$33		& 94	$\pm$24	&7		\\ \hline
\end{tabular}}
\end{center}
\label{dwarf_data}
\end{table*}


\bsp	
\label{lastpage}
\end{document}